\documentclass[aps,pra,twocolumn,groupedaddress,amsmath,amssymb]{revtex4-1}

\usepackage[colorlinks,pdfusetitle,urlcolor=blue,citecolor=blue,linkcolor=blue,bookmarksnumbered,plainpages=false]{hyperref}

\usepackage{color}

\usepackage{bm}
\usepackage{nicefrac}
\usepackage{siunitx}

\usepackage{graphicx}
\usepackage{bm}
\usepackage{braket}

\newcommand{\B}[1]{\mathbf{#1}}

\newcommand{\epsMid}{\epsilon_\text{mid}}
\newcommand{\CasVol}{{\B{V}_{\!\text{C}}}}
\newcommand{\BoxVol}{{\B{V}_\text{B}}}
\newcommand{\para}{\hat{\B{r}}_\parallel  }

\begin{document}

\title{Born-series approach to calculation of Casimir forces}

\author{Robert Bennett}

\affiliation{Department of Physics \& Astronomy, University of Leeds, LS2 9JT, UK}

\date{\today}

\begin{abstract} The Casimir force between two objects is notoriously difficult to calculate in anything other than parallel-plate geometries due to its non-additive nature. This means that for more complicated, realistic geometries one usually has to resort to approaches such as making the crude proximity force approximation (PFA). Another issue with calculation of Casimir forces in real-world situations (such as with realistic materials) is that there are continuing doubts about the status of Lifshitz's original treatment as a true quantum  theory. Here we demonstrate an alternative approach to calculation of Casimir forces for arbitrary geometries which sidesteps both these problems. Our calculations are based upon a Born expansion of the Green's function of the quantised electromagnetic vacuum field, interpreted as multiple scattering, with the relevant coupling strength being the difference in the dielectric functions of the various materials involved. This allows one to consider arbitrary geometries in single or multiple scattering simply by integrating over the desired shape, meaning that extension beyond the PFA is trivial. 
 This work is mostly dedicated to illustration of the method by reproduction of known parallel-slab results -- a process that turns out to be non-trivial and provides several useful insights. We also present a short example of calculation of the Casimir energy for a more complicated geometry, namely that of two finite slabs. 
\end{abstract}

%\keywords{}

\maketitle

%______________________________%
%To do
% Conclusion
% punctuation in equations
%______________________________%

\section{Introduction}\label{Intro}

The existence of an attractive force between two conducting plates, known as the Casimir effect \cite{casimir1948attraction}, is one of the most striking predictions of quantum field theory. Advances in experimental techniques at the nano-scale have meant that the Casimir force has been attracting increasing attention as a phenomenon which may be harnessed in micro and nano-mechanical systems (MEMS/NEMS). Continuing applications include a wide variety of investigations into the forces at work in nanoscale devices relevant to emerging quantum technologies \cite{NetoRoughness, Chan2001MEMS, Munday2006Brownian,capasso2007casimir, rodriguez2011casimir, kats1977VdW,GiesKlingmullerEdgeEffects}. 

These applications have led workers in the field to go far beyond the idealised parallel-plate geometry first considered by Casimir, leading to a number of theories ranging from techniques based directly on Lifshitz's original theory of fluctuations in media \cite{LandauLifshitzPitaevskii, GolestanianCasimir}, to scattering approaches \cite{JaekelReynaudJdP, Rahi2009Scattering}, to geometric optics \cite{jaffe2004casimir} (for comprehensive reviews, see \cite{dalvit2011casimir, bordag2009advances}). These extensions are fraught with difficulties, however, particularly in non-trivial geometries. These problems arise chiefly from the fact that the Casimir force is non-additive, meaning that one cannot reliably approximate the Casimir force for a complex geometry by considering it to be composed of multiple simpler geometries, as is often done across physics and engineering (e.g. finite-element analysis). A scheme known as the proximity force approximation (PFA) attempts to adapt the philosophy of finite-element analysis to Casimir forces, but has been shown many times to be significantly in error when compared to exact results (see, for example, \cite{MaiaNetoPlaneSphere, RodriguezJohnsonPFA}). 

Another contemporary  issue in Casimir physics is much of the literature's reliance upon the theory of intermolecular interactions between extended bodies developed in \cite{LandauLifshitzPitaevskii}, often referred to simply as the Lifshitz theory (see footnote \footnote{The precise definition of what is meant by `Lifshitz theory' has become somewhat ambiguous over time, so much so that almost every Green's function approach to boundary-dependent QED could be classified as such. In this work, any invocation of Lifshitz's name specifically and exclusively refers to the theory presented in \cite{LandauLifshitzPitaevskii}. In this spirit we use the phrase `Lifshitz's original theory' or similar throughout.} concerning terminology). This theory  is the subject of continuing doubts as to its validity as a proper quantum treatment of electrodynamics in media; for example \cite{RosaDalvitMilonniLifshitzClassical} concludes that it is in fact a classical theory. There is no Hamiltonian for the theory presented in \cite{LandauLifshitzPitaevskii}, and it contains no reference to quantised fields. 

An alternative approach to calculation of quantum electrodynamical effects in material bodies is known as macroscopic QED \cite{GrunerWelschOriginalNoiseCurrent} and is based on the introduction of a `noise current' source term in Maxwell's equations in order to satisfy the fluctuation-disspation theorem and preserve the canonical commutation relations of the electromagnetic field operators. It is closely related to the theory of Huttner and Barnett \cite{HuttnerBarnett} where the medium is represented by various interacting matter fields. While extremely elegant and powerful, the model of Huttner and Barnett is somewhat difficult to apply to in homogenous media (see, for example, \cite{EberleinZietalHuttnerBarnett,YeungGustafson}). For this reason we choose to work with the noise-current theory of \cite{GrunerWelschOriginalNoiseCurrent}, which encodes the effects of an inhomogenous medium via its electromagnetic dyadic Green's function -- a quantity which is well-known for several systems of interest. This quality has meant that the noise-current theory has found considerable success in terms of its power and applicability to a wide range of problems (see, for example, \cite{KampfVdW,TomasJPhysA,KhanbekyanFockStatePrep,buhmann2006born}), but in its original form the theory does not attempt a rigorous quantisation (unlike that of Huttner and Barnett), which is a obviously a desirable property for any quantum theory. This suggests that macroscopic QED may have similar problems as Lifshitz's original theory. However, macroscopic QED in the form presented by \cite{GrunerWelschOriginalNoiseCurrent} has relatively recently been put on a firm canonical foundation \cite{PhilbinQuantization, PhilbinCasimir}, in which the noise source term appears naturally. A useful practical consequence of the successful canonical quantisation of \cite{GrunerWelschOriginalNoiseCurrent}'s treatment of macroscopic QED is the elimination of some uncertainties \cite{PhilbinXiongLeonhardt2010} in the correct form of the Casimir energy. This means that for arguably the first time workers in the field are able to confidently use the techniques and results of the noise-current approach to macroscopic QED to investigate the Casimir effect, safe in the knowledge that the theory rests on a canonical foundation.

In this work we will demonstrate an approach to calculating the Casimir force between objects of arbitrary shape that sidesteps the above problems concerning the possible inadequacies of the PFA and Lifshitz's original theory. To do so, we adapt an approach known as dielectric-contrast perturbation theory, where the dyadic Green's function $\boldsymbol{\Gamma}$ that describes the electromagnetic field subject to the boundary conditions imposed by a set of objects is approximated via a Born series \cite{buhmann2006born}. We can then use these approximate Green's functions directly in the formulae presented in \cite{PhilbinCasimir}, meaning our treatment is entirely canonical and independent of Lifshitz's original theory. Furthermore, it will be shown the shape of the objects enters into the calculation as the limits on a volume integral, which can, in principle, be freely chosen. This means that our approach is free from the problems associated with the parallel-plate foundations of the PFA. Arbitrary-order terms in the Born series can be included in a systematic way, this fact turns out to be vital for any calculation of the Casimir effect.

\begin{figure}[h!]
\includegraphics[width = \columnwidth]{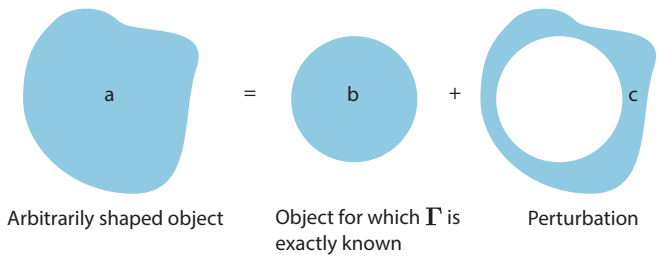}
\caption{(Color online) Illustrative example of the philosophy behind the Born-series expansion approach to finding the electromagnetic dyadic Green's function $\boldsymbol{\Gamma}$ in the vicinity of an object.} \label{GeometricPerturbation}
\end{figure}

The assumption that the Born series approach rests upon is that the dielectric function of an arbitrary arrangement of objects is sufficiently similar to the dielectric function for some simpler arrangement in which we can make exact calculations, as shown schematically in fig.~\ref{GeometricPerturbation}.
We can model the arbitrarily-shaped material body (fig.~\ref{GeometricPerturbation}a) as the sum of a simpler arrangement (\ref{GeometricPerturbation}b) for which $\boldsymbol{\Gamma}$ is known, and a (\ref{GeometricPerturbation}c) for which $\boldsymbol{\Gamma}$ is  not known. The latter can be of arbitrary shape and is considered as a perturbation to the former, and through the machinery of the Born series it can be taken into account to arbitrary precision by including the required number terms. In practice, however, one requires that the `extra' part of the geometry (fig. \ref{GeometricPerturbation}c)  is `small' \footnote{The precise meaning of `small' will be elaborated upon later} in some sense, so that a perturbation series converges quickly. This approach was introduced in \cite{buhmann2006born}, where it was used to calculate Casimir-Polder forces between an atom and a dielectric ring, and also in \cite{GolestanianCasimir}, where the Casimir energy was obtained perturbatively via the original Lifshitz theory for a collection of objects sitting in vacuum. Our work extends and complements that of \cite{GolestanianCasimir} in the following ways:

\begin{enumerate}
\item By far the most significant advance is that this work is based entirely on the canonically-derived Casimir force and energy expressions found in \cite{PhilbinCasimir}. Consequently, this work can be viewed as the first canonical treatment of the dielectric-contrast approach to the Casimir effect, in contrast to \cite{GolestanianCasimir} which is based on the archetypal Lifshitz theory. 
\item Unlike \cite{GolestanianCasimir}, we do not restrict ourselves to vacuum for the medium between the dielectric bodies. This is a significant advance since the theory is based upon there being only a small difference in dielectric function of the collection of material bodies and that of the intervening medium. 
\item  We use a spectral rather than spatial representation of the unperturbed Green's function. While this may seem like a minor technical difference, use of the spectral representation means that our calculation is relatively easily generalizable to more complex unperturbed Green's functions than the homogenous medium we shall use, namely cylinder, sphere and layered versions thereof.  
 \end{enumerate}
 
 In addition to these, a pedagogical difference is that instead of calculating the Casimir energy directly as is done in \cite{GolestanianCasimir}, we systematically piece together a calculation of the Casimir energy, which turns out to mean that we get several useful results and insights along the way.

The structure of this work is as follows. We will begin by reproducing known results for the Casimir force between two parallel, infinite dielectric slabs in section \ref{ForceSection}. We shall see that in order to generalise to finding the force for more complex geometries one actually requires the Casimir energy, so in section \ref{EnergySection} we will also reproduce some known Casimir force results via calculation of the Casimir energy. Finally, in section \ref{NumericalSection} we will present a short numerical example where we investigate the Casimir energy density between finite dielectric slabs.

\section{Casimir Force}\label{ForceSection}
\subsection{Basic expressions}
For dielectric bodies which are translationally invariant in two out of three spatial dimensions, the Casimir force can be expressed entirely in terms of the so-called scattering dyadic Green's function $ \B{G}(\B{r},\B{r}',\omega)$ which describes the geometry-induced modification of the electromagnetic field in the region between the dielectric bodies \cite{RaabeWelschDispersiveForces, PhilbinCasimir}. The $ij$ component of the Casimir stress tensor in Cartesian co-ordinates $\B{r} = \{x,y,z \}$ in a region with position and frequency-dependent permittivity $\epsilon(\B{r},\omega)$ and permeability $\mu(\B{r},\omega)$ \footnote{We work in natural units $\hbar = 1, c=1, \epsilon_0=1$ throughout} is given by an integral over complex frequency $\xi=-i\omega $  \cite{RaabeWelschDispersiveForces, PhilbinCasimir}, namely
\begin{align}
\langle \sigma_{ij}(\B{r}) \rangle =\frac{1}{\pi}\int_0^\infty d\xi & \Big[ \epsilon(\B{r},i\xi) \mathcal{E}_{ij} (\B{r},\B{r},i\xi) \notag \\
&+ \mu^{-1} (\B{r},\B{r},i\xi)\mathcal{B}_{ij} (\B{r},\B{r},i\xi)\Big],\label{BasicForceDefinition}
\end{align}
with the following definitions
\begin{align}
 \mathcal{E}_{ij} (\B{r},\B{r},\omega) &\equiv {\omega^2} \left[ \frac{\delta_{ij}}{2} G^\mathcal{E}_{kk} (\B{r}, \B{r},\omega)- G^\mathcal{E}_{ij} (\B{r}, \B{r},\omega) \right] ,\label{BasicEDef} \\
  \mathcal{B}_{ij} (\B{r},\B{r},\omega) &\equiv \frac{\delta_{ij}}{2} G^\mathcal{B}_{kk} (\B{r}, \B{r},\omega)- G^\mathcal{B}_{ij} (\B{r}, \B{r},\omega) ,\label{BasicBDef}
\end{align}
where 
\begin{align}
\B{G}^\mathcal{E}(\B{r},\B{r},\omega)&\equiv  \B{G}(\B{r},\B{r},\omega), \notag \\
\B{G}^\mathcal{B}(\B{r},\B{r},\omega)&\equiv \lim_{\B{r} \to \B{r}'}  \left[ \nabla \times \B{G}(\B{r},\B{r}',\omega) \times \overleftarrow{\nabla}' \right],\label{GEandGBDefs}
\end{align}
with $ \times \overleftarrow{\nabla}' $ denoting curl with respect to the second index of the tensor $\B{G}_{ij}$, so that $ \B{v}(\B{r}') \times \overleftarrow{\nabla}' \equiv \nabla \times   \B{v}(\B{r}')$ (no minus sign) for a vector $\B{v}(\B{r})$. Eq.~\eqref{BasicForceDefinition} holds for arbitrary media obeying the Kramers-Kronig relations, and its nature as an integral over complex frequencies means it automatically includes the various contributions from different types of mode (travelling, evanescent, surface plasmon, etc, as explicitly shown for related calculations in \cite{bennett2012massshift, bennett2012magmomNJP, bennett2013magmomPRA}). The diagonal components of the Casimir stress tensor directly delivers the force per unit area between two objects, so the problem is essentially reduced to finding the dyadic Green's function of the electromagnetic field for that particular geometry. This is only analytically possible for spheres, infinite planes and cylinders or layered versions thereof since the Helmholtz equation is separable only in those geometries \footnote{Although we remind the reader that Eq.~\eqref{BasicForceDefinition} as shown does not apply for cylinders and spheres}. For more complicated geometries one usually must resort to numerical calculations. 

However, as shown by \cite{buhmann2006born} the dyadic Green's function for an arbitrary geometry may be expanded in a Born (Dyson) series. The defining equation for the \emph{whole} dyadic Green's function $\boldsymbol{\Gamma}(\B{r},\B{r}',\omega)$ (not just its scattering part)  is \footnote{For notional convenience we write the inner product $\B{G}\cdot \B{G}$ of matrices $\B{G}$ as simply $\B{G}\B{G}$, except where doing otherwise would cause ambiguity. The symbol $\times$ placed between two vectors denotes their cross product, and in all other situations is scalar multiplication. The outer product is represented by $\otimes $} 
\begin{align}
\nabla \times &\left[  \mu^{-1}(\B{r},\omega) \nabla \times \boldsymbol{\Gamma}(\B{r},\B{r}',\omega) \right] \notag  \\
&\qquad \qquad -\omega^2 \epsilon(\B{r},\omega)  \boldsymbol{\Gamma}(\B{r},\B{r}',\omega) = \mathbb{I} \delta(\B{r}-\B{r}'),
\end{align}
where $\mathbb{I}$ is the unit dyadic $\mathbb{I}_{ij} = \delta_{ij}$. As shown in \cite{buhmann2006born}, this may be expanded in a Born series about some known `background' Green's function $\boldsymbol{\Gamma}^{(0)}(\B{r},\B{r}',\omega) $
\begin{align}
\boldsymbol{\Gamma}(\B{r}, & \B{r}',\omega) = \boldsymbol{\Gamma}^{(0)}(\B{r},\B{r}',\omega) \notag\\ &+\omega^2 \int d^3 \B{s}_1 \boldsymbol{\Gamma}^{(0)}(\B{r},\B{s}_1,\omega) \delta \epsilon(\B{s}_1,\omega)  \boldsymbol{\Gamma}^{(0)}(\B{s}_1,\B{r}',\omega)\notag \\
&+\omega^4 \int d^3 \B{s}_1\int d^3 \B{s}_2 \Big[ \boldsymbol{\Gamma}^{(0)}(\B{r},\B{s}_1,\omega) \delta \epsilon(\B{s}_1,\omega)\notag\\
& \times \boldsymbol{\Gamma}^{(0)}(\B{s}_1,\B{s}_2,\omega) \delta \epsilon(\B{s}_2,\omega)  \boldsymbol{\Gamma}^{(0)}(\B{s}_2,\B{r}',\omega)\Big]+...
\end{align}
where $\delta \epsilon(\omega)$ is the difference between the entire dielectric function and that of the background material. If we now specify that the system at hand is an object of dielectric function $ \epsilon(\omega)$ described by some volume $\B{V}$ sitting in some `background' dielectric material $\epsilon^{(0)}$ i.e.:
\begin{equation}
 \delta \epsilon(\B{r},\omega)  = \begin{cases} \epsilon(\omega)-\epsilon^{(0)}(\omega)\equiv \delta \epsilon(\omega) \quad &\text{for} \quad  \B{r} \in \B{V}, \\ 0 \quad &\text{for} \quad  \B{r} \notin \B{V}, \end{cases}
\end{equation}
we can restrict the $\B{s}_i$ integrals to being over the volume $\B{V}$, and also bring the dielectric functions outside the integrals
\begin{align}
&\boldsymbol{\Gamma}(\B{r},  \B{r}',\omega) = \boldsymbol{\Gamma}^{(0)}(\B{r},\B{r}',\omega) \notag\\ 
&+\omega^2 [\delta \epsilon(\omega)]  \int_\B{V} d^3 \B{s}_1 \boldsymbol{\Gamma}^{(0)}(\B{r},\B{s}_1,\omega)  \boldsymbol{\Gamma}^{(0)}(\B{s}_1,\B{r}',\omega)\notag \\
&+\omega^4 [\delta \epsilon(\omega) ] ^2 \int_\B{V} d^3 \B{s}_1\int_\B{V} d^3 \B{s}_2 \Big[ \boldsymbol{\Gamma}^{(0)}(\B{r},\B{s}_1,\omega) \notag\\
& \times \boldsymbol{\Gamma}^{(0)}(\B{s}_1,\B{s}_2,\omega)   \boldsymbol{\Gamma}^{(0)}(\B{s}_2,\B{r}',\omega)\Big]+...\notag  \\
=&\, \boldsymbol{\Gamma}^{(0)}(\B{r},\B{r}',\omega) +[\delta \epsilon(\omega)] \boldsymbol{\Gamma}^{(1)}(\B{r},\B{r}',\omega) \notag \\
&\qquad\qquad\qquad\qquad\qquad+[\delta \epsilon(\omega)] ^2\boldsymbol{\Gamma}^{(2)}(\B{r},\B{r}',\omega) +...\notag \\
=&\, \boldsymbol{\Gamma}^{(0)}(\B{r},\B{r}',\omega)+ \sum_{n=1}^\infty [\delta \epsilon(\omega)]^n\boldsymbol{\Gamma}^{(n)}(\B{r},\B{r}',\omega).\label{ScatteringDerivation} \end{align}
The above equation is exact, and holds for any volume $\B{V}$. However, from here on we take the unperturbed Green's function $\boldsymbol{\Gamma}^{(0)}$ to be that for a homogenous medium, which we shall call $\B{H}^{(0)}$;
\begin{equation}
\boldsymbol{\Gamma}^{(0)}(\B{r},  \B{r}',\omega) =\B{H}^{(0)}(\B{r},  \B{r}',\omega) ,
\end{equation}
so that
\begin{align}
\boldsymbol{\Gamma}(\B{r},  \B{r}',\omega) &=\B{H}^{(0)}(\B{r},\B{r}',\omega)+ \sum_{n=1}^\infty \B{H}^{(n)}(\B{r},\B{r}',\omega), \label{WholeGFGeneral} \end{align}
where
\begin{align}
 \sum_{n=1}^\infty & \B{H}^{(n)}(\B{r},\B{r}',\omega)\notag \\
 &=\omega^2 [\delta \epsilon(\omega)]  \int_\B{V} d^3 \B{s}_1 \B{H}^{(0)}(\B{r},\B{s}_1,\omega) \B{H}^{(0)}(\B{s}_1,\B{r}',\omega)\notag \\
&+\omega^4 [\delta \epsilon(\omega) ] ^2 \int_\B{V} d^3 \B{s}_1\int_\B{V} d^3 \B{s}_2 \Big[ \B{H}^{(0)}(\B{r},\B{s}_1,\omega) \notag\\
& \times \B{H}^{(0)}(\B{s}_1,\B{s}_2,\omega) \B{H}^{(0)}(\B{s}_2,\B{r}',\omega)\Big]+...\label{ScatteringDerivationH}
\end{align}
Eq.~\eqref{WholeGFGeneral} describes the whole Green's function. In order to find the scattering Green's function $\B{G}(\B{r},  \B{r}',\omega)$ to be inserted into \eqref{BasicForceDefinition} we must subtract the homogenous part of $\boldsymbol{\Gamma}(\B{r},  \B{r}',\omega) $; 
\begin{align}
\B{G}(\B{r},  \B{r}',\omega) =&\B{H}^{(0)}(\B{r},\B{r}',\omega)+ \sum_{n=1}^\infty \B{H}^{(n)}(\B{r},\B{r}',\omega)\notag \\
&-\B{F}\left[\boldsymbol{\Gamma}(\B{r},\B{r}',\omega)\right],\end{align}
where $\B{F}$ is a functional that extracts the homogenous part of its argument. Using Eq.~\eqref{WholeGFGeneral} we may rewrite this as
\begin{align}
\B{G}(\B{r},  \B{r}',\omega) &=\B{H}^{(0)}(\B{r},\B{r}',\omega)-\B{F}\left[ \B{H}^{(0)}(\B{r},\B{r}',\omega)\right]\notag \\
&+ \sum_{n=1}^\infty \B{H}^{(n)}(\B{r},\B{r}',\omega)- \B{F} \left[\sum_{n=1}^\infty \B{H}^{(n)}(\B{r},\B{r}',\omega) \right]. \label{HomgPartSub}\end{align}
The first line of Eq.~\eqref{HomgPartSub} vanishes by definition, leaving us with
\begin{align}
\B{G}(\B{r},  \B{r}',\omega) &= \sum_{n=1}^\infty[\delta \epsilon(\omega)]^n \left\{\B{H}^{(n)}(\B{r},\B{r}',\omega)- \B{F} \left[\B{H}^{(n)}(\B{r},\B{r}',\omega)\right] \right\}\notag \\
&\equiv  \sum_{n=1}^\infty[\delta \epsilon(\omega)]^n\B{G}^{(n)}(\B{r},  \B{r}',\omega)\, .\label{BasicGFDefinition}\end{align}
We emphasise the fact that $\B{H}^{(n)}(\B{r},\B{r}',\omega)$ has, in general, a homogenous part that must be subtracted, in contrast to refs \cite{GolestanianCasimir} and \cite{buhmann2006born}. As discussed in detail in Appendix \ref{ExampleAppendix}, this complication arises at spatial points where the homogenous part of the dielectric function is not equal to the unperturbed dielectric function. For example, if the unperturbed dielectric function is that for vacuum, and the Green's function is calculated inside an object then the homogenous part of the Green's function is that which would arise if the object were of infinite extent which. This means it is different from the unperturbed Green's function, which would be that for vacuum. This type of situation is not considered in \cite{GolestanianCasimir} or \cite{buhmann2006born}, but its inclusion is necessary here as we shall see in section \ref{EnergySection}. 

\subsection{Calculation of the force}

For our purposes the most convenient form of the homogenous Green's function is a Fourier-space decomposition into vector wave functions \cite{ChewDGFBook}
\begin{align}
&\B{H}(\B{r},\B{r}',\omega ) =-\frac{\hat{\B{z}}\otimes \hat{\B{z}}}{k^2} \delta(\B{r}-\B{r}') + \frac{i}{8\pi^2}\sum_{\B{X}=\B{M},\B{N} }\int d^2\B{k}_\parallel \notag \\
&\times\frac{1}{{k_{z} k_\parallel^2}} \left[\B{X}(\B{k}_\parallel,a k_{z},\B{r})\otimes \B{X}(-\B{k}_\parallel,-a k_{z},\B{r}')\right], \label{HOriginalDef}
\end{align}
with $k_z = \sqrt{k^2-k_\parallel^2}$, $\text{Im} (k_z) >0$ and and $a=\text{sgn}(z-z') $. The functions $\B{M}$ and $\B{N}$ are:
\begin{align}
\B{M}(\B{k}_\parallel, k_z, \B{r})&=i\B{k} \times \hat{\B{z}} \,  e^{i\B{k}\cdot\B{r}} ,\notag \\ \B{N}(\B{k}_\parallel, k_z, \B{r})&=-\frac{1}{k} \B{k}\times \B{k} \times \hat{\B{z}} \, e^{i\B{k}\cdot\B{r}}, \label{MandNDefs}
\end{align}
with $\B{k} = {k}_\parallel \para + k_z \hat{\B{z}}$, where $\para$ is a unit vector perpendicular to $\hat{\B{z}}$. The magnitude $|\B{k}|=k$ of the wave vector $\B{k}$ is given in terms of the frequency $\omega$ by $k^2 = \epsilon(\omega) \mu(\omega) \omega^2 $. The matrix $\hat{\B{z}}\otimes \hat{\B{z}}$ is given by:
\begin{equation}
\hat{\B{z}}\otimes \hat{\B{z}} =\text{diag}(0,0,1)
\end{equation}
The Green's functions for more complicated unperturbed geometries (cylinder, sphere or layered versions thereof) can also be written as products of vector wave-functions in an identical way to \eqref{HOriginalDef}. This is the reason behind our statement in section \ref{Intro} that the spectral form of the Green's function is more appropriate to generalisation of our calculation to more complex unperturbed geometries than the homogenous medium presented here. 

\begin{figure}[h!]
\includegraphics[width = \columnwidth]{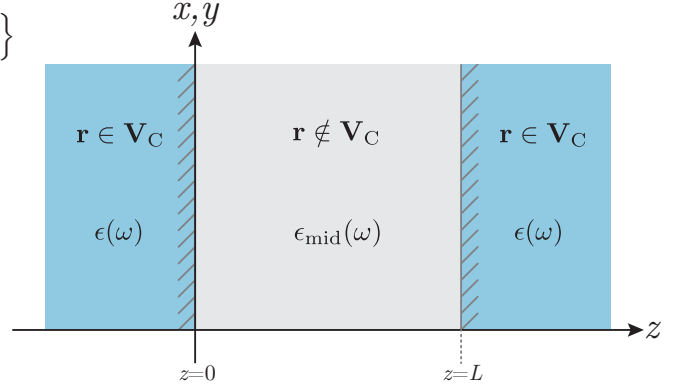}
\caption{(Color online) Three-layer Casimir geometry} \label{PhysicalSetup}
\end{figure}

The Casimir geometry $\CasVol$ we will investigate is shown in Fig.~\ref{PhysicalSetup}.
We will take the unperturbed dielectric function $\epsilon^{(0)}(\omega)$ to be that of the middle slab so that $\epsilon^{(0)}(\omega)=\epsMid(\omega)$. Writing $\B{s} = s_x \hat{\B{x}} +s_y \hat{\B{y}} +s_z \hat{\B{z}}= {s}_\parallel \hat{\B{r}}_\parallel +s_z \hat{\B{z}}$, we have that the Casimir-geometry volume $\CasVol$ over which we shall integrate is given by
\begin{equation}
\int_\CasVol \!\!\! d\B{s} = \int d^2 \B{s}_\parallel \int_{V_z} ds_z=\int d^2 \B{s}_\parallel \left( \int_{-\infty}^0 \!\!\!ds_z + \int_L^\infty \! ds_z \right),
\end{equation}
with $\int d^2 \B{s}_\parallel \equiv \int_{-\infty}^\infty ds_x \int_{-\infty}^\infty ds_y$. To find the force we require the $zz$ component of the Casimir stress tensor \eqref{BasicForceDefinition} in the region between the plates, which depends \emph{only} on the Green's function between the plates, so we may let $\epsilon(\B{r}, \omega)\to \epsMid(\omega)$ in Eq.~\eqref{BasicForceDefinition}, giving:
\begin{align}
\langle \sigma_{ij}(\B{r})  \rangle =\frac{1}{\pi}\int_0^\infty d\xi & \Big[ \epsMid(i\xi) \mathcal{E}_{ij} (\B{r},\B{r},i\xi) + \mathcal{B}_{ij} (\B{r},\B{r},i\xi)\Big] \, . \label{ForcePlates}
\end{align}
We note that $\mathcal{E}_{ij} (\B{r},\B{r},i\xi)$ and $\mathcal{B}_{ij} (\B{r},\B{r},i\xi)$ given by eqs.~\eqref{BasicEDef} and \eqref{BasicBDef} are linear functions of the equal-point scattering Green's function $\B{G}(\B{r},\B{r},i\xi)$, which allows us to define
\begin{align}
 \mathcal{E}_{ij} (\B{r},\B{r},i\xi)&= \sum_{n=1}^\infty[\delta \epsilon(\omega)]^n\mathcal{E}^{(n)}_{ij}(\B{r},  \B{r},\omega),\notag \\
  \mathcal{B}_{ij} (\B{r},\B{r},i\xi)&= \sum_{n=1}^\infty[\delta \epsilon(\omega)]^n\mathcal{B}^{(n)}_{ij}(\B{r},  \B{r},\omega), \label{EandBDefsOrdersStart}
\end{align}
with
\begin{align}
\mathcal{E}^{(n)}_{ij}(\B{r},  \B{r},\omega)&={\omega^2} \left[ \frac{\delta_{ij}}{2} G^{\mathcal{E},{(n)}}_{kk} (\B{r}, \B{r},\omega)- G^{\mathcal{E},{(n)}}_{ij} (\B{r}, \B{r},\omega) \right], \notag \\
\mathcal{B}^{(n)}_{ij}(\B{r},  \B{r},\omega)&=\frac{\delta_{ij}}{2} G^{\mathcal{B},{(n)}}_{kk} (\B{r}, \B{r},\omega)- G^{\mathcal{B},{(n)}}_{ij} (\B{r}, \B{r},\omega),  \label{EandBDefsOrders}
\end{align}
where
\begin{align}
\B{G}^{\mathcal{E},{(n)}}(\B{r},\B{r},\omega)&\equiv  \B{G}^{(n)}(\B{r},\B{r},\omega) ,\notag \\
\B{G}^{\mathcal{B},{(n)}}(\B{r},\B{r},\omega)&\equiv \lim_{\B{r} \to \B{r}'}  \left[ \nabla \times \B{G}^{(n)}(\B{r},\B{r}',\omega) \times \overleftarrow{\nabla}' \right] ,\label{GEandGBDefsOrders}
\end{align}
with $ \B{G}^{(n)}(\B{r},\B{r},\omega)$ given by \eqref{BasicGFDefinition}. Combining definitions \eqref{EandBDefsOrdersStart} to \eqref{GEandGBDefsOrders} gives the following convenient expression for the Casimir stress tensor
\begin{align}
\langle \sigma_{ij} (\B{r}) \rangle =\frac{1}{\pi} \int_0^\infty d\xi \sum_{n=1}^\infty[\delta \epsilon(i\xi)]^n & \Big[ \epsMid(i\xi)\mathcal{E}^{(n)}_{ij} (\B{r},\B{r},i\xi) \notag \\
&+\mathcal{B}^{(n)}_{ij} (\B{r},\B{r},i\xi)\Big] \, .  \label{ForceOrders}
\end{align}
We note that $\mathcal{E}^{(n)}_{ij}$ and $\mathcal{B}^{(n)}_{ij}$ are independent of $\epsilon(\omega)$ so each successive order of approximation of $\langle \sigma_{ij} (\B{r}) \rangle$ about $\epsilon(\omega)=\epsMid(\omega)$ is simply given by plugging each successive term in the Born series of $\mathcal{E}$ and $\mathcal{B}$ into Eq.~\eqref{ForcePlates}. This will not be so simple in the calculation of the Casimir energy due to the requirement to work inside the slabs as well as in the gap between them.

To find the Casimir force between the two slabs we need the Green's function given by Eq.~\eqref{BasicGFDefinition} in the region between the plates only. This region has the special property of the unperturbed Green's function being equal to the homogenous part of the whole Green's function. As shown in appendix \ref{ExampleAppendix}, this means that the following must hold;
\begin{equation}
 \B{F} \left[\B{H}^{(n)}(\B{r},\B{r},\omega)\right]=0 \qquad \text{for }\B{r} \notin \CasVol ,
\end{equation}
leaving us with
\begin{align}
\B{G}_\text{mid}(\B{r},  \B{r},\omega) &= \sum_{n=1}^\infty[\delta \epsilon(\omega)]^n \B{H}^{(n)}(\B{r},\B{r},\omega) ,\label{GBetweenPlatesGeneral}  \end{align}
where we have used the obvious notation 
\begin{equation}
\B{G}_\text{mid}(\B{r},  \B{r},\omega)=\B{G}(\B{r},  \B{r},\omega) \qquad \text{for }\B{r} \notin \CasVol \, . 
\end{equation}
Equation \eqref{GBetweenPlatesGeneral} makes physical sense because if the unperturbed Green's function is equal to the homogenous part of the whole Green's function then the scattering Green's function coincides with the perturbation to that homogenous part. Again we emphasize that this is $\emph{not}$ true in general, it is specifically not true in a region where the unperturbed Green's function is different from the homogenous part of the whole Green's function, as discussed in detail in appendix \ref{ExampleAppendix}.

\subsubsection{First order}\label{ForceFirstOrderSection}

\begin{figure}[h!]
\includegraphics[width = 0.6\columnwidth]{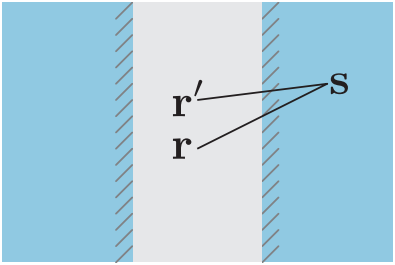}
\caption{(Color online) A term with a single scattering event $\B{s}$ cannot contribute to the Casimir force.} \label{SDFirstOrder}
\end{figure}

To find the contribution $\B{G}^{(1)}_\text{mid}(\B{r},  \B{r},\omega)=\B{G}^{(1)}(\B{r} \notin \CasVol,\B{r} \notin \CasVol, \omega)$ to the Green's function between the slabs that is first-order in $\delta \epsilon(\omega)$ we need to evaluate the $n=1$ term of Eq.~\eqref{GBetweenPlatesGeneral}. This is given by
\begin{align}
\B{G}^{(1)}_\text{mid}&(\B{r},  \B{r},\omega) =\omega^2 [\delta \epsilon(\omega)] \notag \\
&\times  \int_\CasVol d^3 \B{s}_1 \B{H}^{(0)}(\B{r},\B{s}_1,\omega) \B{H}^{(0)}(\B{s}_1,\B{r},\omega) \, .  \label{G1MidExplicit}
\end{align}
Since $\B{s}_1$ runs only over the volume $\CasVol$ and we are calculating $\B{G}^{(1)}$ in the region $\B{r} \notin \CasVol$, we may immediately ignore the $\delta$ function part of $\B{H}$ shown in Eq.~\eqref{HOriginalDef}. Denoting the parallel frequency integration variable contained within one of the factors $\B{H}^{(0)}$ in the integrand of Eq.~\eqref{G1MidExplicit} as $\B{p}_\parallel$, one easily sees that the $\B{s}_\parallel$ integral trivially evaluates to $\delta$ functions over $\B{k}_\parallel$ and $\B{p}_\parallel$, meaning that the $\B{p}_\parallel$ integral is also trivial. Going into polar co-ordinates defined by $k_x =k_\parallel \cos \theta$ and $k_y=k_\parallel \sin \theta$ and carrying out the angular integral $\int_0^{2\pi}d\theta $, we find for the first-order terms $\B{G}^{\mathcal{E},(1)}$ and $\B{G}^{\mathcal{B},(1)}$ in the approximations of $\B{G}^\mathcal{E}$ and $\B{G}^\mathcal{B}$ [Eqs.~\eqref{GEandGBDefsOrders}]:
\begin{align}
\B{G}^{\mathcal{X},(1)}(\B{r} \notin \CasVol, &\B{r} \notin \CasVol   ,i\xi)\equiv \B{G}_{\text{mid}}^{\mathcal{X},(1)}(\B{r},\B{r} ,i\xi)\notag \\
=&-\frac{1}{16 \pi \epsMid(\omega)^2\xi^2 }\int_{V_z} ds_{1z} \int_0^\infty dk_\parallel k_\parallel  \notag \\
& \times  \frac{m^{(1)}_\mathcal{X}(k_\parallel, \xi)}{\kappa^2(k_\parallel,\xi)}e^{-2 \left|z-s_{1z}\right| \kappa(k_\parallel,\xi)}, \label{GFFirstOrderCenter}
\end{align}
with $\mathcal{X} = \mathcal{E}, \mathcal{B}$, and where we have defined
\begin{equation}
\kappa(k_\parallel,\xi)=\sqrt{\epsMid(i\xi) \xi ^2+k_\parallel^2}\, . 
\end{equation}
The quantities $m^{(1)}_\mathcal{X}$ are $3 \times 3$ matrices given by
\begin{equation}
m^{(1)}_\mathcal{X}(k_\parallel, \xi)=\text{diag}(m^{(1)}_{\mathcal{X}, xx},m^{(1)}_{\mathcal{X},yy},m^{(1)}_{\mathcal{X},zz}) ,
\end{equation}
with
\begin{align}
m^{(1)}_{\mathcal{E}, xx}&=2 k_\parallel^4+3\epsMid(i\xi) k_\parallel^2 \xi^2  +2 \epsMid^2(i\xi) \xi ^4=m^{(1)}_{\mathcal{E},yy}, \notag \\
m^{(1)}_{\mathcal{E}, zz} &=2k_\parallel^2 (2k_\parallel^2 +\epsMid(i\xi) \xi^2), \notag\\
m^{(1)}_{\mathcal{B}, xx}&=\xi ^4 \epsMid^2(i\xi) [3 k_\parallel^2+2 \xi ^2 \epsMid(i\xi) ]=m^{(1)}_{\mathcal{B}, yy},\notag\\
m^{(1)}_{\mathcal{B}, zz} &=2 \xi ^4 k_\parallel^2 \epsMid^2(i\xi)\, .  \label{FirstOrderMiddleMatrixElements}
\end{align}
The choice of notation $m^{(1)}_\mathcal{X}$ reminds the reader that these quantities are specific to the \emph{m}iddle of the geometry (the region between the slabs), later on we shall consider more general versions of these matrix elements. We note that the matrix elements $m_{\mathcal{B},ij}^{(1)}$ pertaining to the magnetic-type terms are not obtainable from the electric-type matrix elements  $m_{\mathcal{E},ij}^{(1)}$ through a multiplicative factor, as one might expect from a duality relation $\epsilon \leftrightarrow \mu$ \cite{BuhmannScheelDuality}. The reasons for this are discussed in appendix \ref{DualityAppendix}. The $s_{1z}$ integral in Eq.~\eqref{GFFirstOrderCenter} is elementary, 
\begin{align}
&\left( \int_{-\infty}^0ds_{1z}+\int_L^\infty ds_{1z} \right)  e^{-2\kappa(k_\parallel,\xi) \left|z-s_{1z}\right| }\notag \\
&=\frac{1}{{2\kappa(k_\parallel,\xi)}}\left(  e^{-2 z\kappa(k_\parallel,\xi)}+e^{2 (z-L)\kappa(k_\parallel,\xi)}\right) \, . 
\end{align}
Using this in Eq.~\eqref{GFFirstOrderCenter}, we find agreement with the linear term in the Taylor expansion of the exact Green's function \cite{Tomas2002} for $\epsilon(\omega) \approx \epsilon_\text{mid}(\omega)$ with $\B{r}$ between the slabs \footnote{Care must be taken when comparing to ref.~\cite{Tomas2002} due to a factor $4\pi$ difference in the definition of the Green's function}. Combining Eqs.~\eqref{EandBDefsOrders}-\eqref{ForceOrders} and \eqref{GFFirstOrderCenter}, it is easy to see that the integrand for the first order term in $zz$ component of the stress tensor $\langle \sigma_{ij} (\B{r}) \rangle$ must be proportional to
\begin{align}
   -\xi^2\epsMid(\omega) &\left[ m^{(1)}_{\mathcal{E}, xx}+m^{(1)}_{\mathcal{E}, yy}-m^{(1)}_{\mathcal{E}, zz} \right] \notag \\
& +\left[ m^{(1)}_{\mathcal{B}, xx}+m^{(1)}_{\mathcal{B}, yy}-m^{(1)}_{\mathcal{B}, zz} \right]=0, \label{ForceProportionalityFirstOrder}
\end{align}
so we conclude that the Casimir stress tensor (and consequently the force) vanishes to linear order in the dielectric contrast $\epsilon(\omega)-\epsMid(\omega)$;
\begin{equation}
\langle \sigma_{zz} (\B{r}) \rangle^{(1)} = 0 \; . 
\end{equation}
This result makes physical sense because the first term in the Born series only `knows' about one plate since it contains only a single scattering event, as shown in fig.~\ref{SDFirstOrder}. This means there cannot be a Casimir force to this order since the attractive force between the objects results from their interaction with each other, mediated by the electromagnetic field. Clearly this interaction cannot take place if there is only one scattering event. 

\subsubsection{Second order} \label{SecondOrderForceSection}

\begin{figure}[h!]
\includegraphics[width = 0.6\columnwidth]{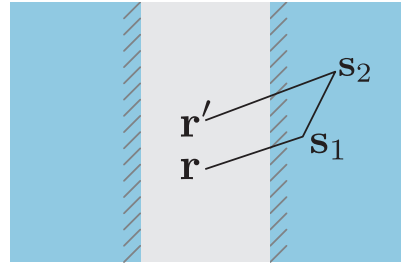}
\caption{(Color online) A term with two scattering events $\B{s}_1$ and $\B{s}_2$ in the same slab as each other cannot contribute to the Casimir force. The Case shown corresponds to the terms labelled as `RR' in the text.} \label{SecondOrderSameSlabFig}
\end{figure}

\begin{figure}[h!]
\includegraphics[width = 0.6\columnwidth]{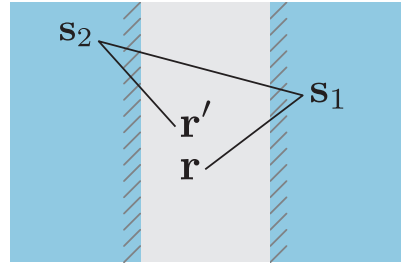}
\caption{(Color online) A term with a two scattering events $\B{s}_1$ and $\B{s}_2$ in different slabs \emph{can} contribute to the Casimir force. The case shown corresponds to terms labelled as `RL' in the text.} \label{SecondOrderDiffSlabFig}
\end{figure}

The second-order Green's function in the region between the plates is given by the $n=2$ term of Eq.~\eqref{GBetweenPlatesGeneral}.
\begin{align}
&\B{G}^{(2)}_\text{mid}(\B{r},\B{r},\omega) =\omega^4  [\delta \epsilon(\omega) ] ^2 \int_\CasVol d^3 \B{s}_1\int_\CasVol d^3 \B{s}_2  \notag\\
& \times\Big[ \B{H}^{(0)}(\B{r},\B{s}_1,\omega)  \B{H}^{(0)}(\B{s}_1,\B{s}_2,\omega)   \B{H}^{(0)}(\B{s}_2,\B{r},\omega)\Big] \label{GSecondOrderBetweenPlates}
\end{align}
As explained below Eq.~\eqref{G1MidExplicit}, in the first-order calculation the restriction $\B{r} \notin \CasVol$, equivalent to $\B{r} \neq \B{s}_1$, meant we could ignore the delta function part of the homogenous Green's functions $\B{H}^{(0)}(\B{r},\B{s}_1,\omega)$ and $\B{H}^{(0)}(\B{s}_1,\B{r},\omega)$ entering into Eq.~\eqref{G1MidExplicit}. However, in the second-order calculation we have an additional factor of $\B{H}^{(0)}(\B{s}_1,\B{s}_2,\omega) $, meaning we have a term proportional to $\delta(\B{s}_1-\B{s}_2)$. This does not vanish since the $\B{s}_1$ and $\B{s}_2$ integrals run over the same region as each other. Because of this we rewrite the homogenous Green's function as the sum of a delta function part $-({\hat{\B{z}}\otimes \hat{\B{z}}}/{k^2} )\delta(\B{r}-\B{r}') $ (contributing at equal spatial points only), and a `propagating' part $ \B{H}^{(0)}_\text{prop}(\B{r},\B{r}',\omega ) $ which contributes for all spatial points:
\begin{align}
&\B{H}^{(0)}(\B{r},\B{r}',\omega ) =-\frac{\hat{\B{z}}\otimes \hat{\B{z}}}{k^2} \delta(\B{r}-\B{r}') + \B{H}^{(0)}_\text{prop}(\B{r},\B{r}',\omega )\; , \label{HSplit} \end{align}
with
\begin{align}
& \B{H}^{(0)}_\text{prop}(\B{r},\B{r}',\omega ) = \frac{i}{8\pi^2}\sum_{\B{X}=\B{M},\B{N} }\int d^2\B{k}_\parallel \notag \\
&\times\frac{1}{{k_{z} k_\parallel^2}} \left[\B{X}(\B{k}_\parallel,a k_{z},\B{r})\otimes \B{X}(-\B{k}_\parallel,-a k_{z},\B{r}')\right] \, . \label{HPropDefinition}
\end{align}
Since $\B{r}\notin \CasVol $, we have for the two outer factors in the second line of \eqref{GSecondOrderBetweenPlates}
\begin{align}
\B{H}^{(0)}(\B{r},\B{s}_1,\omega)  =\B{H}^{(0)}_\text{prop}(\B{r},\B{s}_1,\omega ) ,\\
\B{H}^{(0)}(\B{s}_2,\B{r},\omega)  =\B{H}^{(0)}_\text{prop}(\B{s}_2,\B{r},\omega ) ,
\end{align}
but for the middle factor:
\begin{equation}
\B{H}^{(0)}(\B{s}_1,\B{s}_2,\omega)  =-\frac{\hat{\B{z}}\otimes \hat{\B{z}}}{k^2} \delta(\B{s}_1-\B{s}_2) +\B{H}^{(0)}_\text{prop}(\B{s}_1,\B{s}_2,\omega ) \, .
\end{equation}
This means the second-order equal-point Green's function naturally splits into two parts,
\begin{equation}
\B{G}^{(2)}_\text{mid}(\B{r},\B{r},\omega)  = \B{G}^{(2)}_\text{prop}(\B{r},\B{r},\omega) + \B{G}^{(2)}_\delta(\B{r},\B{r},\omega),
\end{equation}
where
\begin{align}
&\B{G}^{(2)}_\text{prop}(\B{r},\B{r},\omega) =\omega^4  [\delta \epsilon(\omega) ] ^2 \int_\CasVol d^3 \B{s}_1\int_\CasVol  d^3 \B{s}_2,  \notag\\
& \times\Big[ \B{H}^{(0)}_\text{prop}(\B{r},\B{s}_1,\omega)  \B{H}^{(0)}_\text{prop}(\B{s}_1,\B{s}_2,\omega)   \B{H}^{(0)}_\text{prop}(\B{s}_2,\B{r},\omega)\Big] ,\label{GPropCenter}
\end{align}
and 
\begin{align}
\B{G}^{(2)}_\delta(\B{r},\B{r},\omega) &=-\omega^4  [\delta \epsilon(\omega) ] ^2\frac{1}{k^2} \int_\CasVol d^3 \B{s}_1\int_\CasVol  d^3 \B{s}_2   \notag\\
& \!\!\!\!\!\!\!\!\!\!\!\!\!\!\!\!\!\!\!\!\times\Big[ \B{H}^{(0)}_\text{prop}(\B{r},\B{s}_1,\omega) \hat{\B{z}}\otimes \hat{\B{z}}\delta(\B{s}_1-\B{s}_2)   \B{H}^{(0)}_\text{prop}(\B{s}_2,\B{r},\omega)\Big],\notag \\
&=-\omega^4  [\delta \epsilon(\omega) ] ^2\frac{1}{k^2}  \int_\CasVol d^3 \B{s}_1 \notag\\
& \!\!\!\!\!\!\!\!\!\!\!\!\!\!\!\!\!\!\!\!\times\Big[ \B{H}^{(0)}_\text{prop}(\B{r},\B{s}_1,\omega)  (\hat{\B{z}}\otimes \hat{\B{z}})\B{H}^{(0)}_\text{prop}(\B{s}_1,\B{r},\omega)\Big], \label{GExtraForce}
\end{align}
with $k^2 = \epsMid(\omega) \omega^2$. It should be noted that Eq.~\eqref{GExtraForce} only makes reference to a single intermediate point $\B{s}_1$, so it is not expected to contribute to the Casimir force. In component form, we have for Eq.~\eqref{GExtraForce}
\begin{align}
\B{G}^{(2)}_{\delta,ij}(\B{r},\B{r},\omega)&=-\omega^4  [\delta \epsilon(\omega) ] ^2\frac{1}{k^2}  \int_\CasVol d^3 \B{s}_1 \notag\\
& \!\!\!\!\!\!\!\!\!\!\!\!\!\!\!\!\!\!\!\!\times\Big[ \B{H}^{(0)}_{\text{prop},iz}(\B{r},\B{s}_1,\omega)  \B{H}^{(0)}_{\text{prop},zj}(\B{s}_1,\B{r},\omega)\Big] \, . \label{GDeltaFinal}
\end{align}
Though we shall not repeat the algebra here, the above analysis for also holds for $\nabla \times \B{G}^{(2)}_\text{mid}(\B{r},\B{r},\omega) \times \overleftarrow{\nabla}'$ since the plane-wave nature of the homogenous Green's function means that the derivative operators can only generate various overall factors of $k_x, k_y$ or $k_z$. Proceeding, we substitute Eq.~\eqref{HPropDefinition} into \eqref{GDeltaFinal}, make the same integral manipulations that took Eq.~\eqref{G1MidExplicit} to \eqref{GFFirstOrderCenter} and evaluate the integral over $\B{s}_1$, giving
\begin{align}
\B{G}^{\mathcal{X},{(2)}}_{\delta,ij}&(\B{r},\B{r},\omega)=\frac{1}{32 \pi  \xi ^2\epsMid^3(i\xi)} \int_0^\infty \frac{dk_\parallel k_\parallel^3 }{\kappa(k_\parallel,\xi)}\notag \\
&\times m^{(1)}_{\delta,\mathcal{X}}(k_\parallel,\xi)   \left[e^{-2z \kappa(k_\parallel,\xi)}+e^{2 (z-L) \kappa(k_\parallel,\xi)}\right] \; , 
\end{align}
where $\mathcal{X} = \mathcal{E}, \mathcal{B}$, and
\begin{align}
\B{G}^{ \mathcal{E},{(2)}}_{\delta,ij}(\B{r},\B{r},\omega)&\equiv  \B{G}^{{(2)}}_{\delta,ij}(\B{r},\B{r},\omega), \\
\B{G}^{\mathcal{B}, {(2)} }_{\delta,ij}(\B{r},\B{r},\omega)&\equiv \lim_{\B{r}\to \B{r}'} \left[ \nabla \times \B{G}^{{(2)}}_{\delta,ij}(\B{r},\B{r}',\omega) \times \overleftarrow{\nabla}' \right],
\end{align}
[c.f. Eqs.~\eqref{EandBDefsOrders}]. The matrices $m^{(1)}_{\delta, \mathcal{X}}$ are defined as:
\begin{equation}
m^{(1)}_{\delta, \mathcal{X}}(k_\parallel, \xi)=\text{diag}(m^{(1)}_{\delta, \mathcal{X}, xx},m^{(1)}_{\delta, \mathcal{X},yy},m^{(1)}_{\delta, \mathcal{X},zz}) ,
\end{equation}
with elements 
\begin{align}
m^{(1)}_{\delta, \mathcal{E}, xx}&=1=m^{(1)}_{\delta, \mathcal{E},yy}, \notag \\
m^{(1)}_{\delta, \mathcal{E}, zz} &=\frac{2 k_\parallel ^2}{k_\parallel^2+\xi ^2 \epsMid(i\xi)},\notag\\
m^{(1)}_{\delta, \mathcal{B}, xx}&=\frac{\xi ^4 \epsMid^2(k_\parallel,\xi)}{k_\parallel^2+\xi ^2  \epsMid(i\xi)}=m^{(1)}_{\delta, \mathcal{B}, yy},\notag\\
m^{(1)}_{\delta, \mathcal{B}, zz} &=0\, . \label{FirstOrderMiddleMatrixElementsDelta}
\end{align}
In an identical way to section \ref{ForceFirstOrderSection}, we then note that the Casimir force arising from this term is proportional to
\begin{align}
-\xi^2\epsMid(\omega) &\left[ m^{(1)}_{\delta, \mathcal{E}, xx}+m^{(1)}_{\delta, \mathcal{E}, yy}-m^{(1)}_{\delta, \mathcal{E}, zz} \right]\notag \\
&+\left[ m^{(1)}_{\delta, \mathcal{B}, xx}+m^{(1)}_{\delta, \mathcal{B}, yy}-m^{(1)}_{\delta, \mathcal{B}, zz} \right]=0\, . 
\end{align}
So, as expected, the single-scattering part of the second-order Green's function [given by Eq.~\eqref{GExtraForce}] does not contribute to the Casimir force. We are then left with the double-scattering part only, given by Eq.~\eqref{GPropCenter}. The full integral obtained by substituting Eq.~\eqref{HPropDefinition} into \eqref{GPropCenter}  is far too cumbersome to write explicitly here, but it is worth noting that its integrand is proportional to
\begin{align}
&\text{exp}\bigg\{-\kappa(k_\parallel,\xi) \Big[ (z-s_{1z}) \text{sgn}(z-s_{1z})\notag \\
&+z \, \text{sgn}(z-s_{2z})+s_{1z} \text{sgn}(s_{1z}-s_{2z})\notag \\
&-s_{2z} \text{sgn}(s_{1z}-s_{2z})+s_{2z} \text{sgn}(s_{2z}-z)\Big]\bigg\},
\end{align}
which results in different behaviour depending on the relative signs of $z$, $s_{1z}$ and $s_{2z}$, in contrast to the first-order calculation. To best interpret this behaviour we first note that for the Casimir geometry $\CasVol$ one has
\begin{align}
\int_\CasVol \!\!\! d^3 \B{s}_1\int_\CasVol \!\!\! d^3 \B{s}_2 =& \int d^2 \B{s}_{1\parallel}\int d^2  \B{s}_{2\parallel}\left[ \int_{-\infty}^0 \!\!\! ds_{1z}+\int_{L}^\infty \!\!\! ds_{1z}\right]\notag \\ &\times \left[ \int_{-\infty}^0 ds_{2z}+\int_{L}^\infty ds_{2z}\right],
\end{align}
with $\int d^2 \B{s}_{i \parallel} \equiv \int_{-\infty}^\infty ds_{ix} \int_{-\infty}^\infty ds_{iy}$. This means that we can split up each contribution into those from two scattering events in one slab (either the left or right slab, denoted by LL or RR), and those from one scattering in each slab (right-to-left or left-to-right, denoted by RL and LR), i.e.:
\begin{align}
\B{G}^{(2)}_\text{prop}(\B{r},\B{r}',\omega)=&\B{G}^{(2)}_\text{LL}(\B{r},\B{r}',\omega)+\B{G}^{(2)}_\text{RR}(\B{r},\B{r}',\omega)\notag \\
&+\B{G}^{(2)}_\text{LR}(\B{r},\B{r}',\omega)+\B{G}^{(2)}_\text{RL}(\B{r},\B{r}',\omega), \label{BasicSOGFBetweenPlates}
\end{align}
where, for example, the spatial integral for the LR term is over the region $\int_{-\infty}^0 ds_{1z}\int_{L}^\infty ds_{2z} $. Since the Casimir force \eqref{ForcePlates} is linear in the scattering Green's function we can individually assign contributions from each of the four terms in Eq.~\eqref{BasicSOGFBetweenPlates}.
\begin{equation}
\langle \sigma_{zz} \rangle^{(2)}= \langle \sigma_{zz} \rangle^{(2)}_{\text{LL}} +\langle \sigma_{zz} \rangle^{(2)}_{\text{RR}} +\langle \sigma_{zz} \rangle^{(2)}_{\text{LR}} +\langle \sigma_{zz} \rangle^{(2)}_{\text{RL}}, \label{ForceFourTerms}
\end{equation}
where, for example, $\langle \sigma_{zz} \rangle^{(2)}_{\text{RR}}$ is found from Eqs.~\eqref{BasicForceDefinition}-\eqref{GEandGBDefs} by taking $\B{G}\to \B{G}^{(2)}_\text{RR}$.  Carrying out a considerable amount of algebra, again making the same integral manipulations that took Eq.~\eqref{G1MidExplicit} to \eqref{GFFirstOrderCenter} and writing our result in the same form as Eq.~\eqref{GDeltaFinal}, we find:
\begin{align}
\B{G}_{\text{LL}}^{\mathcal{X},(2)}(\B{r},& \B{r},i\xi)=\frac{[\delta\epsilon(i\xi)]^2}{64 \pi [\epsMid(i\xi)]^2 } \int_0^\infty dk_\parallel k_\parallel\notag \\
& \times \frac{ m^{(2)}_{\mathcal{X},\text{LL}}(k_\parallel, \xi)}{ (\epsMid(i\xi) \xi ^2+k_\parallel^2)^{5/2}} e^{-2(L-z) \kappa(k_\parallel,\xi)},
\end{align}
with the equivalent for the other slab being obtained by a reflection of the coordinate system and translation by a distance $L$;
\begin{equation}
\B{G}_{\text{RR}}^{\mathcal{X},(2)}=\B{G}_{\text{LL}}^{\mathcal{X},(2)}(z \to L-z)\, .  \label{LeftToRightTrans}
\end{equation}
Similarly
\begin{align}
&\B{G}_{\text{LR}}^{\mathcal{X},(2)}(\B{r},\B{r}, i\xi)=\frac{[\delta\epsilon(\omega)]^2}{128 \pi [\epsMid(\omega)]^3 } \int_0^\infty dk_\parallel k_\parallel \notag \\
&\!\!\!\!\!\times \frac{ m^{(2)}_{\mathcal{X},\text{LR}}(k_\parallel, \xi)}{ (\epsMid(\omega) \xi ^2+k_\parallel^2)^{5/2}} e^{-2L \kappa(k_\parallel,\xi)}=\B{G}_{\text{RL}}^{\mathcal{X},(2)}(\B{r},\B{r},i\xi) ,\label{GSecondOrderForceLR}
\end{align}
where, anticipating the known result \cite{SchwingerCasimir, Tomas2002}, we note that the integrand is constant in $z$. The matrices $m^{(2)}_{\mathcal{X},\lambda}$ are given by
\begin{equation}
m^{(2)}_{\mathcal{X},\lambda}(k_\parallel, \xi)=\text{diag}(m^{(2)}_{\mathcal{X},\lambda, xx},m^{(2)}_{\mathcal{X},\lambda,yy},m^{(2)}_{\mathcal{X},\lambda,zz}) ,
\end{equation}
with $\lambda = \text{LL},\text{LR}$ and $\mathcal{X} = \mathcal{E}, \mathcal{B}$. The LL matrix elements coincide with those from the first-order calculation \footnote{One should not attach too much physical meaning to this fact, it is simply a consequence of the way $\B{G}_{\mathcal{X}}^{(1)}$ and $\B{G}_{\mathcal{X},\text{LR}}^{(2)}$ were defined} 
\begin{align}
m^{(2)}_{\mathcal{X},\text{LL}}=m^{(1)}_{\mathcal{X}}\; .  \label{SecondvsFirstOrderM}
\end{align}
Combining Eqs.~\eqref{BasicSOGFBetweenPlates}-\eqref{SecondvsFirstOrderM}, one eventually finds agreement with the quadratic term in the Taylor expansion of the exact Green's function \cite{Tomas2002} for $\epsilon(\omega) \approx \epsilon_\text{mid}(\omega)$ with $\B{r}$ between the slabs. This means that our results for the Casimir force are necessarily going to  agree with those obtained using the exact Green's function, so it seems that our investigation of the Born series method is complete. However, it turns out that considering the terms in \eqref{BasicSOGFBetweenPlates} individually provides some physical insight, so we proceed to work out the Casimir force using our approximations. Once again using identical reasoning to that shown in section \ref{ForceFirstOrderSection}, we note that contribution from the LL terms to the Casimir force integrand is proportional to
\begin{align}
-\xi^2\epsMid(i\xi) &\left[ m^{(2)}_{ \mathcal{E}, LL, xx}+m^{(2)}_{\mathcal{E},  LL, yy}-m^{(2)}_{\mathcal{E},  LL, zz} \right]\notag \\
&+\left[ m^{(2)}_{\mathcal{B},  LL, xx}+m^{(2)}_{\mathcal{B},  LL, yy}-m^{(2)}_{\mathcal{B},  LL, zz} \right], \label{LLCombinationForce}
\end{align}
Combining Eqs.~\eqref{ForceProportionalityFirstOrder}, \eqref{LeftToRightTrans}, \eqref{SecondvsFirstOrderM} and \eqref{LLCombinationForce}, we immediately see that the contribution from the LL and RR terms both vanish;
\begin{equation}
\langle \sigma_{zz} \rangle^{(2)}_{\text{LL}} = 0=\langle \sigma_{zz} \rangle^{(2)}_{\text{RR}} \label{LLandRRVanish}.
\end{equation} 
This is expected because the LL and RR terms represent the contribution of terms with two scattering events in one slab [as shown in Fig.~\eqref{SecondOrderSameSlabFig}],  
so the term does not represent interaction between the slabs, meaning that no Casimir force can arise from it. Moving on to the LR and RL contributions, 
the matrix elements entering into \eqref{GSecondOrderForceLR} are
\begin{align}
m^{(2)}_{\mathcal{E}, \text{LR}, xx}&=4 k_\parallel^6+8 k_\parallel^4 \epsMid(i\xi) \xi ^2+5 k_\parallel^2 \epsMid^2(i\xi) \xi ^4\notag \\
&\qquad \qquad\qquad\qquad+2 \epsMid^3(i\xi) \xi ^6=m^{(2)}_{\mathcal{E},\text{LR}, yy} \notag ,\\
m^{(2)}_{\mathcal{E}, \text{LR}, zz} &=-2k_\parallel^2 \left[2 k_\parallel^2+ \epsMid(i\xi) \xi ^2\right]^2,\notag\\
m^{(2)}_{\mathcal{B}, \text{LR}, xx}&=-\epsMid^2(i\xi) \xi ^4 \Big[ 4 k_\parallel^4+5 k_\parallel^2 \epsMid(i\xi) \xi ^2\notag \\
&\qquad \qquad\qquad\qquad+2 \epsMid^2(i\xi) \xi ^4\Big]=m^{(2)}_{\mathcal{B}, \text{LR}, yy},\notag\\
m^{(2)}_{\mathcal{B}, \text{LR}, zz} &=2 k_\parallel^2 \epsMid^3(i\xi) \xi ^6. 
\end{align}
Combining Eqs.~\eqref{ForceFourTerms}, \eqref{GSecondOrderForceLR} and \eqref{LLandRRVanish}, we find for the second order contribution to the Casimir force
\begin{align}
\langle \sigma_{zz} \rangle^{(2)} &=- \frac{1}{16 \pi ^2 }\int_0^\infty d\xi\frac{[\epsilon(i\xi)-\epsMid(i\xi)]^2}{\epsMid^2(i\xi)}\int_0^\infty dk_\parallel   k_\parallel\notag \\
&\times\frac{e^{-2 L\kappa(k_\parallel,\xi)}}{\kappa^3(k_\parallel,\xi)} \left[2 k_\parallel^4+2 k_\parallel^2 \epsMid(i \xi) \xi ^2+\epsMid^2(i \xi) \xi ^4\right] \notag  \\
&\qquad\qquad\qquad+\mathcal{O}\left\{[\epsilon(i\xi)-\epsMid(i\xi)]^3\right\} .\label{SigmaZZSecondOrderFinal}
\end{align}
This does not vanish because the scattering events at $\B{s}_1$ and $\B{s}_2$ were in different slabs, as shown in Fig.~\eqref{SecondOrderDiffSlabFig}. 
For non-dispersive slabs $\epsMid(i \xi) \to \epsMid$ and $\epsilon(i \xi)\to \epsilon$ the double integral in \eqref{SigmaZZSecondOrderFinal} becomes elementary. The result is;
\begin{equation}
\langle \sigma_{zz} \rangle^{(2)} \!=\!-\frac{23}{640 \pi^2 L^4}\frac{(\epsilon-\epsMid)^2}{ \epsMid^{5/2}}+\mathcal{O}\left[(\epsilon-\epsMid)^3\right], \label{ForceSecondOrderFinalResult}
\end{equation}
which reduces to the corresponding result of \cite{SchwingerCasimir} for $\epsMid=1$. It is worth noting that for dispersive slabs the dielectric function may in practice take on a wide range of values as the frequency is integrated over -- in a more elaborate calculation one would have to be very careful that no particular value of $\xi$ causes the Born series to diverge. Finally we note that the integral formula \eqref{SigmaZZSecondOrderFinal} can be reproduced by Taylor expansion of the well-known Lifshitz formula for the Casimir force.

\section{Casimir Energy}\label{EnergySection}
\subsection{Basic expressions}

The Casimir force expression \eqref{BasicForceDefinition} only holds in systems that are translationally invariant in two out of three spatial directions, i.e. infinite parallel slabs or layered variants thereof. Since the utility of the method used here is in calculation of the Casimir force between objects of arbitrary shape, we need to consider an alternative approach, more general approach to calculating Casimir forces in the system. This approach consists of finding the Casimir energy density throughout the system, then determining the force from this by varying the distance between the objects. In this section we demonstrate how to calculate the Casimir energy density in the same geometry as in section \ref{ForceSection}. This situation was investigated in order to facilitate a comparison with known results, reproduction of which turns out to be non-trivial due to several pitfalls associated with using dielectric-contrast perturbation theory to find the Green's function in a region whose dielectric function is different from the unperturbed dielectric function (a case which is not present in \cite{GolestanianCasimir, buhmann2006born} or section \ref{ForceSection} of this work). These complications would not be immediately obvious in a direct brute-force numerical application of dielectric-contrast method to find the Casimir force between two objects with some complex geometry. 

The Casimir energy density $\langle \rho (\B{r}) \rangle$ is given in Cartesian co-ordinates by \cite{PhilbinCasimir}:
\begin{align}
\langle \rho (\B{r}) \rangle =&\frac{1}{2\pi}\int_0^\infty d\xi  \Big[-\xi^2 \frac{d[\xi \epsilon(\B{r},i\xi)]}{d\xi} \text{tr} \B{G}^\mathcal{E}(\B{r},\B{r},i\xi) \notag \\
&\quad \,+\frac{1}{\mu^2(\B{r},i\xi)}\frac{d[\xi \mu(\B{r},i\xi)]}{d\xi} \text{tr}  \B{G}^\mathcal{B}(\B{r},\B{r},i\xi)\Big] \, ,\label{PhilbinOriginalEnergy}
\end{align}
with $\B{G}^\mathcal{E}$ and $\B{G}^\mathcal{B}$ given by Eq.~\eqref{GEandGBDefs}. This means that for the Casimir geometry shown in fig.~\ref{PhysicalSetup} we have %
\begin{align}
\langle \rho (\B{r}) \rangle =\frac{1}{2\pi} \int_0^\infty d\xi  \Big[-\xi^2 \epsilon(z,i\xi) & \text{tr} \B{G}^\mathcal{E}(\B{r},\B{r},\omega) \notag  \\
&+\text{tr}  \B{G}^\mathcal{B}(\B{r},\B{r},\omega)\Big], \label{EnergyStartPoint}
\end{align}
with
\begin{equation}
\epsilon(z,\omega) = \begin{cases} \epsilon(\omega) \qquad &\text{for } z<0 \text{ or } z>L,\\
\epsilon_\text{mid}(\omega) \qquad &\text{for } 0<z<L. \end{cases}
\end{equation}
In order to find the Casimir force from the Casimir energy density one needs to find the total Casimir energy in the system, then differentiate with respect to the distance between the slabs. This means that, in contrast to what we needed for the direct Casimir force expression, we need to find the Green's function inside each slab, as well as in the gap between them. Another consequence of this is that we need to take into account the complication briefly touched upon in section \ref{ForceSection} just after Eq.~\eqref{ForceOrders}. To see how this happens we use Eqs.~\eqref{GEandGBDefsOrders} to write Eq.~\eqref{EnergyStartPoint} for $\B{r}\in \CasVol$ as
\begin{align}
&\langle \rho(\B{r}\in \CasVol) \rangle =\frac{1}{2\pi}\lim_{\B{r} \to \B{r}'} \int_0^\infty d\xi  \notag \\
&\!\!\times \!\!\Bigg\{ \delta\epsilon(i\xi)  \text{tr} \Big[-\xi^2 \epsilon(i \xi) \B{G}^{\mathcal{E},{(1)}}(\B{r},\B{r},i \xi) +\B{G}^{\mathcal{B},{(1)}}(\B{r},\B{r},i\xi)\Big]\notag \\
&\qquad\qquad\qquad+[\delta \epsilon(i\xi) ]^2\text{tr} \Big[-\xi^2 \epsilon(i \xi) \B{G}^{\mathcal{E},{(2)}}(\B{r},\B{r},i\xi)\notag  \\
 &\qquad\qquad\qquad\qquad\qquad+ \B{G}^{\mathcal{B},{(2)}}(\B{r},\B{r},i\xi)\Big]+...\Bigg\} \; .
\end{align}
It is tempting to say that the two terms in the above equation represent the contributions  to the Casimir energy of first and second-order in $\delta \epsilon(i\xi)$, but since $\B{r} \in \CasVol$ this is \emph{not} the case. This subtle point arises because the dielectric function that appears in the energy expression in the coefficient of $\B{G}^{\mathcal{E}}(\B{r},\B{r},\omega)$ is $\epsilon(\omega)$, not $\epsilon_\text{mid}(\omega)$. Noting that the following relation holds
\begin{align}
\epsilon(\omega) [\delta \epsilon(\omega)] = \epsMid(\omega) \left[\epsilon(\omega)-\epsMid(\omega)\right]&\notag \\
&\!\!\!\!\!\!\!\!\!\!\!\!\!\!\!\!\!\!\!\!\!\!\!\!\!\!\!\!\!\!\!\!\!\!\!\!+ \left[\epsilon(\omega)-\epsMid(\omega)\right]^2 ,
\end{align}
we have the contribution to the energy density for $\B{r}\in \CasVol$ that is linear in $\delta \epsilon(\omega)$:
\begin{align}
&\langle  \rho(\B{r}\in \CasVol)  \rangle^{(1)}  =\frac{1}{2\pi}\int_0^\infty d\xi  \delta \epsilon(i\xi) \notag \\
& \times  \text{tr} \Big[-\xi^2 \epsMid(i\xi)  \B{G}^{(1)}_\mathcal{E}(\B{r},\B{r},i\xi)+  \B{G}^{(1)}_\mathcal{B}(\B{r},\B{r},i\xi)\Big],\label{FirstOrderEnergyInSlabs}
\end{align}
and the contribution that is quadratic in $\delta \epsilon(\omega)$:
\begin{align}
&\langle \rho(\B{r}\in \CasVol) \rangle ^{(2)} =\frac{1}{2\pi}\int_0^\infty d\xi [\delta \epsilon(i\xi)]^2 \notag \\
& \times \text{tr} \Big\{-\xi^2\left[ \B{G}^{(1)}_\mathcal{E}(\B{r},\B{r},i\xi)+ \epsMid(i\xi)  \B{G}^{(2)}_\mathcal{E}(\B{r},\B{r},\omega)\right]  \notag \\
&\qquad \qquad\qquad \qquad\qquad \qquad +  \B{G}^{(2)}_\mathcal{B}(\B{r},\B{r},i\xi)\Big\}\; .  \label{SecondOrderEnergyInSlabs}
\end{align}
This demonstrates the unexpected result that a first-order Green's function appears in the second-order term in the approximation of the Casimir energy density in the region $\B{r} \in \CasVol$. If we were to take this calculation to third order, a second-order Green's function would appear in the third-order Casimir energy and so on. For $\B{r} \notin \CasVol$ there is no such complication and we have simply
\begin{align}
\langle  \rho&(\B{r}\notin \CasVol)  \rangle^{(1)}  =\frac{1}{2\pi} \int_0^\infty d\xi  \delta \epsilon(i\xi) \notag \\
& \times  \text{tr} \Big[-\xi^2 \epsMid(i\xi)  \B{G}^{(1)}_\mathcal{E}(\B{r},\B{r},i\xi)+  \B{G}^{(1)}_\mathcal{B}(\B{r},\B{r},i\xi)\Big],\label{FirstOrderEnergyInCenter}
\end{align}
and
\begin{align}
&\langle \rho(\B{r}\notin \CasVol) \rangle ^{(2)} =\frac{1}{2\pi}\int_0^\infty d\xi [\delta \epsilon(i\xi)]^2 \notag \\
& \times \text{tr} \Big\{-\xi^2\epsMid(i\xi)  \B{G}^{(2)}_\mathcal{E}(\B{r},\B{r},i\xi) + \B{G}^{(2)}_\mathcal{B}(\B{r},\B{r},i\xi)\Big\}. \label{SecondOrderEnergyInCenter}
\end{align}

\subsection{Calculation of Casimir Energy}

\subsubsection{First order}\label{FirstOrderEnergySection}

We begin by using Eq.~\eqref{HSplit} to expand the $n=1$ contribution to the first term of the second line of Eq.~\eqref{BasicGFDefinition} for distinct $\B{r}$ and $\B{r}'$. The result, before subtraction of the homogenous part [c.f. Eq.~\eqref{BasicGFDefinition}] is
%\B{G}_{ij}^{(1)}(\B{r},\B{r}',\omega)&=\omega^2 [\delta \epsilon(\omega)]  \int_\B{V} d^3 \B{s}_1 \left[ - \frac{\B{z}\otimes \B{z}}{k^2}\delta(\B{r}-\B{s}_1)+\B{H}^{(0)}_\text{prop}(\B{r},\B{s}_1,\omega) \right]_{ik} \left[ - \frac{\B{z}\otimes \B{z}}{k^2}\delta(\B{s}_1-\B{r}')+\B{H}^{(0)}_\text{prop}(\B{s}_1,\B{r}',\omega) \right]_{kj} \notag \\
%
\begin{align}
&{\B{G}}_{ij}^{(1)}(\B{r},\B{r}',\omega)=\omega^2 [\delta \epsilon(\omega)] \notag\\
&  \times \Bigg[ \int_\CasVol d^3 \B{s}_1\B{H}^{(0)}_{\text{prop},ik}(\B{r},\B{s}_1,\omega)\B{H}^{(0)}_{\text{prop},kj}(\B{s}_1,\B{r}',\omega )\notag\\
&- \frac{\delta_{iz}}{k^2}\B{H}^{(0)}_{\text{prop},zj}(\B{r}\in\CasVol,\B{r}',\omega)  - \frac{\delta_{jz}}{k^2}\B{H}^{(0)}_{\text{prop},iz}(\B{r},\B{r}'\in \CasVol,\omega)\notag \\
&\quad+\frac{\delta_{iz}\delta_{jz}}{k^4}\begin{cases} \delta(\B{r}-\B{r}') &\text{ if } \B{r},\B{r}' \in \CasVol \\
0 &\text{ otherwise}\end{cases}\Bigg]-\B{F},\label{FirstOrderEnergy}\end{align}
where we have carried out $\B{s}_1$ integrals over $\delta$ functions where possible. In order to avoid a proliferation of notation it is understood that $\B{F}$ without argument represents the homogenous part of the expression preceding it, and we have adopted a notation for $\B{H}$ that is particularly useful in later calculations:
\begin{equation}
\B{H}(\B{r} \in \B{V}, \B{r}' \in \B{V}',\omega) \equiv \begin{cases} \B{H}(\B{r} , \B{r}' ,\omega) &\text{if }  \B{r} \in \B{V}, \\
\B{H}(\B{r} , \B{r}' ,\omega) &\text{if }  \B{r}' \in \B{V}',\\
0 &\text{otherwise}. \end{cases}
\end{equation}
The spatial dependence of $\B{H}^{(0)}_{\text{prop}}$ is given by the difference $\B{r}-\B{r}'$, not their absolute values. This means the second and third terms of Eq.~\eqref{FirstOrderEnergy} are position-independent at $\B{r} = \B{r}'$, so vanish upon subtraction of a homogenous part once the limit $\B{r}\to \B{r}'$ is taken. The same is also easily seen to apply to the final term of Eq.~\eqref{FirstOrderEnergy}, meaning that we are left with its first term only, minus its homogenous part. 
\begin{align}
&\B{G}_{ij}^{(1)}(\B{r},\B{r},\omega)=\omega^2 [\delta \epsilon(\omega)] \notag \\
& \times \left[  \int_\CasVol d^3 \B{s}_1 \B{H}^{(0)}_{\text{prop},ik}(\B{r},\B{s}_1,\omega)\B{H}^{(0)}_{\text{prop},kj}(\B{s}_1,\B{r},\omega )\right]-\B{F},\label{GijFirstOrderMiddle}\end{align}
As discussed in section \ref{ForceSection}, one must be very careful when working in a region where the dielectric function and the unperturbed dielectric function are not equal -- in particular we shall see that the first term of \eqref{FirstOrderEnergy} for $\B{r} \in \CasVol$ does indeed have a  homogenous part which needs to be subtracted.

\begin{figure}[h!]
\includegraphics[width=\columnwidth]{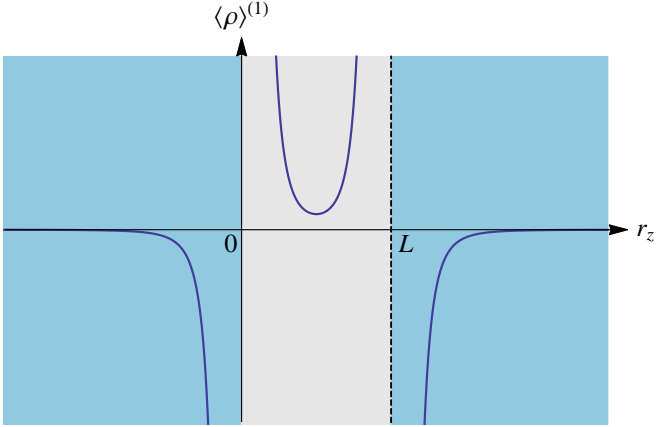}
\caption{(Color online) First-order approximation to the energy density in the three regions. The energy density diverges on the plates, but the total energy is finite, and in this case is zero, leading to zero Casimir force as demonstrated in section \ref{ForceFirstOrderSection}.} \label{EnergyDensityGraph}
\end{figure}

For $\B{r} \notin \CasVol$, Eq.~\eqref{GijFirstOrderMiddle} coincides with the $\B{r} \to \B{r}'$ limit of the calculation shown in section \ref{ForceFirstOrderSection}, with the final result being given by Eq.~\eqref{GFFirstOrderCenter}. However here we need $\B{r}\in \CasVol$ as well. Carrying out the calculation of integrals \eqref{G1MidExplicit} for general $\B{r}$, we find for the generalization $\B{G}_{\text{gen}}^{\mathcal{X},(1)}(\B{r},\B{r},i\xi)$ of $\B{G}_{\text{mid}}^{\mathcal{X},(1)}(\B{r},\B{r},i\xi)$ [Eq.~\eqref{GFFirstOrderCenter}]
\begin{align}
&\B{G}_{\text{gen}}^{\mathcal{X},{(1)}}(\B{r},\B{r},i\xi)=-\frac{\delta\epsilon(i\xi)}{16 \pi \epsMid^2(i\xi)\xi^2 }\bigg[ \int_{V_z} ds_{1z} \int_0^\infty dk_\parallel \notag \\
& \times \frac{k_\parallel g^{(1)}_\mathcal{X}(k_\parallel, \xi)}{ \epsMid(i\xi) \xi ^2+k_\parallel^2} e^{-2 \left|z-s_{1z}\right| \kappa(k_\parallel,\xi)} -\B{F}\bigg]\label{GX1Gen},\end{align}
with
\begin{align}
&g^{(1)}_{\mathcal{E},xx}=\epsMid^2(i\xi) \xi ^4+\bigg\{ \left[k_\parallel^2+\epsMid(i\xi) \xi ^2\right] \text{sgn}[z-s_z]^2\notag \\
& \times \left[k_\parallel^2+\text{sgn}[z-s_z]^2(k_\parallel^2+\epsMid(i\xi) \xi ^2) \right] \bigg\}=g^{(1)}_{\mathcal{X},yy},\notag \\
&g^{(1)}_{\mathcal{E},zz}=2k_\parallel^2 \left\{k_\parallel^2+\left[k_\parallel^2+\epsMid(i\xi) \xi ^2\right] \text{sgn}[z-s_z]^2\right\},\notag\\
&g^{(1)}_{\mathcal{B},xx}=-k_\parallel^6- \bigg\{ \left[k_\parallel^2+\xi ^2 \epsMid(i\xi)\right]^2 \text{sgn}[z-s_z]^2 \notag \\
&\times \Big[k_\parallel^2 \text{sgn}[z-s_z]^2-\left[k_\parallel^2+\xi ^2 \epsMid(i\xi)\right] \text{sgn}[z-s_z]^2 \notag\\
&\qquad \qquad \qquad\qquad   +k_\parallel^2-\xi ^2 \epsMid(i\xi)\Big]\bigg\}=g^{(1)}_{\mathcal{B},yy}\notag\\
&g^{(1)}_{\mathcal{B},zz}=2 \xi ^4 k_\parallel^2  \epsMid(i\xi) ,
\end{align}
which are seen to reduce to Eqs.~\eqref{FirstOrderMiddleMatrixElements} when $z \neq s_z$ (i.e. $\B{r} \notin \CasVol$);
\begin{equation}
g^{(1)}_\mathcal{X}(k_\parallel, \xi)|_{z \neq s_z}=m^{(1)}_\mathcal{X}(k_\parallel, \xi). \label{FirstOrderMatrixElementsGtoM}
\end{equation}
 Carrying out the $s_z$ integral in Eq.~\eqref{GX1Gen}, one finds a homogenous part stemming from the point $z \neq s_z$. Since we subtract this we can use Eq.~\eqref{FirstOrderMatrixElementsGtoM} to write for the left slab ($z <0$) 
\begin{align}
\B{G}^{(1),\mathcal{X}}&(z<0,i\xi)=-\frac{\delta\epsilon(i\xi)}{32 \pi \epsMid^2(i\xi)\xi^2 }\int_0^\infty  \frac{dk_\parallel k_\parallel}{\kappa^3(k_\parallel,\xi)} \notag  \\ 
&\times {m^{(1)}_\mathcal{X}(k_\parallel, \xi)} \left[  {e^{2 (z-L)\kappa(k_\parallel,\xi)}-e^{2 z\kappa(k_\parallel,\xi)}} \right],\label{G1ZLessThan}\end{align}
with the corresponding quantity for the right slab being obtained as usual by taking $z \to L-z$. For the central region we find
\begin{align}
\B{G}^{(1),\mathcal{X}}&(0<z<L,i\xi)=-\frac{\delta\epsilon(i\xi)}{32 \pi \epsMid^2(i\xi)\xi^2 }\int_0^\infty  \frac{dk_\parallel k_\parallel}{\kappa^3(k_\parallel,\xi)} \notag \\
&\times m^{(1)}_\mathcal{X}(k_\parallel, \xi)\left[  {e^{2 (z-L)\kappa(k_\parallel,\xi)}+e^{-2 z\kappa(k_\parallel,\xi)}} \right],\label{G1ZCentral}\end{align}
which is in agreement with \cite{Tomas2002}. We are now in a position to calculate energy density to first-order in $\delta \epsilon(\omega)$ in all three regions of the Casimir system shown in fig.~\ref{PhysicalSetup}. Beginning in the left slab $z <0$, we have from Eqs.~\eqref{FirstOrderEnergyInSlabs} and \eqref{G1ZLessThan}
\begin{align}
\langle \rho  \rangle^{(1)}&(z<0) =- \frac{1}{64 \pi^2}\int_0^\infty \frac{d\xi }{\xi^2 }\frac{\delta\epsilon(i\xi)}{ \epsMid^2(i\xi)}\int_0^\infty \frac{k_\parallel dk_\parallel }{\kappa^3 (k_\parallel,\xi)}\notag \\
& \times \text{tr}  \left[-\xi^2  \epsMid(i\xi)m^{(1)}_\mathcal{E}(k_\parallel, \xi)  + m^{(1)}_\mathcal{B}(k_\parallel, \xi)\right]\notag\\ & \times  \left[  {e^{2 (z-L)\kappa(k_\parallel,\xi)}-e^{2 z\kappa(k_\parallel,\xi)}} \right] ,\label{FirstOrderEnergyDensityLeft}  
\end{align}
with the corresponding quantity for the right hand slab being obtained via
\begin{equation}
\langle \rho  \rangle^{(1)}(z>L)=\left[ \langle \rho  \rangle^{(1)}(z<0)\right] \bigg|_{z\to L-z},\label{FirstOrderEnergyDensityRight}  
\end{equation}
In the central region we have from Eqs.~\eqref{FirstOrderEnergyInSlabs} and \eqref{G1ZCentral}
\begin{align}
\langle \rho \rangle^{(1)}(0& <z<L) =-\frac{1}{64\pi^2}\int_0^\infty \frac{d\xi }{\xi^2 }\frac{\delta\epsilon(i\xi)}{ \epsMid^2(i\xi)} \int_0^\infty \frac{k_\parallel dk_\parallel }{\kappa^3(k_\parallel,\xi)}\notag \\
& \times \text{tr}   \left[-\xi^2  \epsMid(i\xi)m^{(1)}_\mathcal{E}(k_\parallel, \xi)  +  m^{(1)}_\mathcal{B}(k_\parallel, \xi)\right]\notag\\ & \times  \left[  {e^{2 (z-L)\kappa(k_\parallel,\xi)}+e^{-2 z\kappa(k_\parallel,\xi)}} \right]\; . \label{FirstOrderEnergyDensityCenter}  
\end{align}
A plot of the qualitative behaviour of eqs.~\eqref{FirstOrderEnergyDensityLeft} and \eqref{FirstOrderEnergyDensityCenter} in the three regions of the Casimir geometry $\CasVol$ is shown in Fig.~\ref{EnergyDensityGraph}.  
  The total energy is given by integrating each of these over the spatial region for which each is valid. The first two lines of eqs.~\eqref{FirstOrderEnergyDensityLeft} and \eqref{FirstOrderEnergyDensityCenter} are the same as each other and independent of $z$, so, in combination with Eq.~\eqref{FirstOrderEnergyDensityRight}, it is easy to see that the total energy in the system is proportional to
\begin{align}
&\int_{-\infty}^0 dz  \left[  {e^{2 (z-L)\kappa(k_\parallel,\xi)}-e^{2 z\kappa(k_\parallel,\xi)}} \right] \notag \\
+&\int_0^L dz\left[  {e^{2 (z-L)\kappa(k_\parallel,\xi)}+e^{-2 z\kappa(k_\parallel,\xi)}} \right] \notag\\
+&\int_{L}^\infty dz  \left[  {e^{-2z\kappa(k_\parallel,\xi)}-e^{2 (L-z)\kappa(k_\parallel,\xi)}} \right] . \label{FirstOrderEnergyDensityIntegrals}
\end{align}
Changing variables to $z \to L-z$ in the final term transforms it to being identical to the first term, which is expected since the total energy contained in the left slab should be equal to the total energy contained in the right slab. Evaluating the integrals we find that the quantity shown in Eq.~\eqref{FirstOrderEnergyDensityIntegrals} vanishes, meaning that there is no Casimir energy in the system to linear order in $\epsilon(\omega)-\epsMid(\omega)$. This of course says nothing about the total electromagnetic energy in the system, since homogenous parts were subtracted throughout. These would give a non-zero total electromagnetic energy, but that is not of interest here. This result is of course in trivial agreement with the results of section \ref{ForceFirstOrderSection}, where we found zero Casimir force to first order in $\delta\epsilon(\omega)$.

%The traces are:
%
%\begin{align}
%\text{tr}  \left[m^{(1)}_\mathcal{E}(k_\parallel, \xi)\right]&=4\xi^2 \epsMid(\omega) \left(2k_\parallel^4/(\xi^2 \epsMid(\omega) ) +2 k_\parallel^2 + \xi^2 \epsMid(\omega) \right)\notag \\
%\text{tr}  \left[m^{(1)}_\mathcal{B}(k_\parallel, \xi)\right]&=4 \xi ^4\epsMid(\omega)^2 \left(2 k_\parallel^2+\xi ^2\epsMid(\omega) \right)
%\end{align}
%
%giving
%
%\begin{align}
%\langle \rho \rangle^{(1)} =&\frac{\delta\epsilon(\omega)}{8 \epsMid(\omega) \pi^2} \int_0^\infty d\xi \int_0^\infty dk_\parallel \frac{ k_\parallel^5 }{{( \epsMid(\omega) \xi ^2+k_\parallel^2)^{3/2}}}\notag\\ & \times  \left[  {e^{2 (z-d)\kappa(k_\parallel,\xi)}-e^{2 z\kappa(k_\parallel,\xi)}} \right]  
%\end{align}

\subsubsection{Second order}
\begin{figure}[h!]
\includegraphics[width = 0.6\columnwidth]{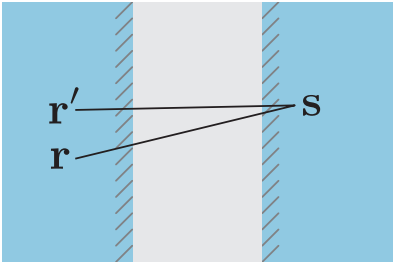}
\caption{(Color online) First order (single scattering event) terms \emph{can} contribute to the Casimir energy inside the slabs.} \label{SDFirstOrderInSlabFig}
\end{figure}

The second order scattering Green's function inside the slabs is given by the $n=2$ term of Eq.~\eqref{BasicGFDefinition}
\begin{align}
&\B{G}^{(2)}(\B{r},\B{r}',\omega) =\omega^4  [\delta \epsilon(\omega) ] ^2 \int_\CasVol d^3 \B{s}_1\int_\CasVol d^3 \B{s}_2  \notag\\
&\!\!\!\times\Big[ \B{H}^{(0)}(\B{r},\B{s}_1,\omega)  \B{H}^{(0)}(\B{s}_1,\B{s}_2,\omega)   \B{H}^{(0)}(\B{s}_2,\B{r}',\omega)\Big] -\B{F}. \label{GSecondOrderEnergyStart}
\end{align}
The $ij$ component of the expression in square brackets in Eq.~\eqref{GSecondOrderEnergyStart} is:
\begin{align}
&\left[ - \frac{\delta_{iz}\delta_{kz}}{k^2}\delta(\B{r}-\B{s}_1)+\B{H}^{(0)}_{\text{prop},ik}(\B{r},\B{s}_1,\omega) \right] \notag \\ &\quad \times \!\left[ - \frac{\delta_{lz}\delta_{kz}}{k^2}\delta(\B{s}_1-\B{s}_2)+\B{H}^{(0)}_{\text{prop},kl}(\B{s}_1,\B{s}_2,\omega) \right]\notag \\ &\qquad \quad  \times\!\left[ - \frac{{\delta_{lz}\delta_{jz}}}{k^2}\delta(\B{s}_2-\B{r}')+\B{H}^{(0)}_{\text{prop},lj}(\B{s}_2,\B{r}',\omega) \right] . \label{ijExpansion}
\end{align}
We will sort the terms of \eqref{ijExpansion} by the number of $\delta$ functions appearing in each. We have the single term with three $\delta$ functions
\begin{align}
&\B{G}^{(2)}_{ij,\delta\delta\delta}(\B{r},\B{r}',\omega) =- \frac{\delta_{iz}\delta_{jz}}{k^6}\int_\CasVol d^3 \B{s}_1\int_\CasVol d^3 \B{s}_2 \delta(\B{r}-\B{s}_1)\notag \\ &\qquad \qquad\qquad \qquad\times \delta(\B{s}_1-\B{s}_2)\delta(\B{s}_2-\B{r}')\notag \\
&= - \frac{\delta_{iz}\delta_{jz}}{k^6}\begin{cases} \delta(\B{r}-\B{r}') &\text{ if }   (\B{r}\in \CasVol) \land  (\B{r}'\in \CasVol)\\
0 &\text{otherwise}\end{cases} \label{TripleDelta}.
\end{align}
We will require the scattering Green's function at $\B{r}=\B{r}'$, and we also need to subtract any part which does not depend on $\B{r}$. Thus, Eq.~\eqref{TripleDelta} cannot contribute to the scattering Green's function. Now consider the terms with two $\delta$ functions
%
%\begin{align}
%\B{G}^{(2)}_{ij,\delta \delta}(\B{r},\B{r}',\omega) & =\frac{\omega^4  }{k^4} [\delta \epsilon(\omega) ] ^2 \int_\B{V} d^3 \B{s}_1\int_\B{V} d^3 \B{s}_2\bigg[ \delta_{iz}\delta_{kz}\delta_{kz}\delta_{lz}\delta(\B{r}-\B{s}_1) \delta(\B{s}_1-\B{s}_2)\B{H}^{(0)}_{\text{prop},lj}(\B{s}_2,\B{r}',\omega)\notag \\
%  & \qquad  +\delta_{iz}\delta_{kz}\delta_{jz}\delta_{lz}\delta(\B{r}-\B{s}_1) \B{H}^{(0)}_{\text{prop},kl}(\B{s}_1,\B{s}_2,\omega)\delta(\B{s}_2-\B{r}')+\delta_{lz}\delta_{kz}\delta_{jz}\delta_{lz}\B{H}^{(0)}_{\text{prop},ik}(\B{s}_1,\B{s}_2,\omega)\delta(\B{s}_1-\B{s}_2) \delta(\B{s}_2-\B{r}')\bigg]\notag\\
% &\!\!\!\!\!\!\!\!\!\!\!\!\!\!\!\!\!\!\!\!=\frac{\omega^4}{k^4}  [\delta \epsilon(\omega) ] ^2 \left[ {\delta_{iz}}\B{H}^{(0)}_{\text{prop},zj}(\B{r}\in \B{V},\B{r}',\omega)+ {\delta_{iz}\delta_{jz}}\B{H}^{(0)}_{\text{prop},zz}(\B{r}\in \B{V},\B{r}'\in \B{V},\omega)+{\delta_{jz}}\B{H}^{(0)}_{\text{prop},iz}(\B{r},\B{r}'\in \B{V},\omega)\right]
%\end{align}
%
\begin{align}
\B{G}^{(2)}_{ij,\delta \delta}(\B{r},\B{r}',\omega) =& \,\frac{\omega^4}{k^4}  [\delta \epsilon(\omega) ] ^2 \bigg[ {\delta_{iz}}\B{H}^{(0)}_{\text{prop},zj}(\B{r}\in \CasVol,\B{r}',\omega)\notag \\
&\quad+ {\delta_{iz}\delta_{jz}}\B{H}^{(0)}_{\text{prop},zz}(\B{r}\in \CasVol,\B{r}'\in \CasVol,\omega)\notag \\
&\quad\quad+{\delta_{jz}}\B{H}^{(0)}_{\text{prop},iz}(\B{r},\B{r}'\in\CasVol ,\omega)\bigg] .\label{GSecondOrderTwoDeltas}
\end{align}
As discussed in section \ref{FirstOrderEnergySection}, $\B{H}^{(0)}_{\text{prop},zj}$ depends only on the difference $\B{r}-\B{r}'$, so at $\B{r}=\B{r}'$ the terms in Eq.~\eqref{GSecondOrderTwoDeltas} must all be independent of $\B{r}$, so cannot contribute to the scattering Green's function. Next we look at the terms in the expansion of of \eqref{ijExpansion} that contain one $\delta$ function:
%
%\begin{align}
%\B{G}^{(2)}_{ij,\delta}(\B{r},\B{r}',\omega) &=-\frac{\omega^4  }{k^2} [\delta \epsilon(\omega) ] ^2\int_\B{V} d^3 \B{s}_1\int_\B{V} d^3 \B{s}_2  \Big[  \delta_{iz}\delta_{kz} \delta(\B{r}-\B{s}_1)\B{H}^{(0)}_{\text{prop},kl}(\B{s}_1,\B{s}_2,\omega)\B{H}^{(0)}_{\text{prop},lj}(\B{s}_2,\B{r}',\omega) \notag \\
%& \qquad+ \delta_{lz}\delta_{kz}\B{H}^{(0)}_{\text{prop},ik}(\B{r},\B{s}_1,\omega) \delta(\B{s}_1-\B{s}_2)\B{H}^{(0)}_{\text{prop},lj}(\B{s}_2,\B{r}',\omega)+\delta_{lz}\delta_{jz}\B{H}^{(0)}_{\text{prop},ik}(\B{r},\B{s}_1,\omega) \B{H}^{(0)}_{\text{prop},kl}(\B{s}_1,\B{s}_2,\omega)\delta(\B{s}_2-\B{r}')\Big]\notag  \\
%&=-\frac{1}{k^2}\int_\B{V} d^3 \B{s}   \Big[ \delta_{iz} \B{H}^{(0)}_{\text{prop},zl}(\B{r}\in \B{V},\B{s},\omega)\B{H}^{(0)}_{\text{prop},lj}(\B{s},\B{r}',\omega)+\B{H}^{(0)}_{\text{prop},iz}(\B{r},\B{s},\omega) \B{H}^{(0)}_{\text{prop},zj}(\B{s},\B{r}',\omega)\notag  \\
%&\qquad +\delta_{jz}\B{H}^{(0)}_{\text{prop},ik}(\B{r},\B{s},\omega) \B{H}^{(0)}_{\text{prop},kz}(\B{s},\B{r}'\in \B{V},\omega)\Big]
%\end{align}
\begin{align}
& \B{G}^{(2)}_{ij,\delta}(\B{r},\B{r}',\omega) =-\frac{\omega^4  [\delta \epsilon(\omega) ] ^2 }{k^2}\int_\CasVol d^3 \B{s} \notag   \\& \times \Big[ \delta_{iz} \B{H}^{(0)}_{\text{prop},zl}(\B{r}\in \CasVol,\B{s},\omega)\B{H}^{(0)}_{\text{prop},lj}(\B{s},\B{r}',\omega)\notag \\
&\quad  +\B{H}^{(0)}_{\text{prop},iz}(\B{r},\B{s},\omega) \B{H}^{(0)}_{\text{prop},zj}(\B{s},\B{r}',\omega)\notag  \\
&\quad \quad +\delta_{jz}\B{H}^{(0)}_{\text{prop},ik}(\B{r},\B{s},\omega) \B{H}^{(0)}_{\text{prop},kz}(\B{s},\B{r}'\in \CasVol,\omega)\Big].
\end{align}
These terms contribute to the scattering Green's function since, in general, $\B{s} \neq \B{r}, \B{r}'$. In contrast to the calculation for the force, terms with a single intermediate scattering point $\B{s}$ \emph{can} contribute to the Casimir energy since the point $\B{r}$ may be inside one of the slabs meaning that if $\B{s}$ is in the other slab, as shown Fig.~\ref{SDFirstOrderInSlabFig},
there is a term that `knows' about both slabs. Finally we have the term without $\delta$ functions
\begin{align}
&\B{G}^{(2)}_{ij,0}(\B{r},\B{r}',\omega) =\omega^4  [\delta \epsilon(\omega) ] ^2 \int_\CasVol d^3 \B{s}_1\int_\CasVol d^3 \B{s}_2  \notag  \\
&\times \B{H}^{(0)}_{\text{prop},ik}(\B{r},\B{s}_1,\omega)  \B{H}^{(0)}_{\text{prop},kl}(\B{s}_1,\B{s}_2,\omega) \B{H}^{(0)}_{\text{prop},lj}(\B{s}_2,\B{r}',\omega),
\end{align}
so that upon restoration of the subtraction of the homogenous part, the entire second-order contribution to the scattering Green's function at $\B{r}=\B{r}'$ is given by
\begin{equation}
\B{G}^{(2)}(\B{r},\B{r},\omega) =\B{G}^{(2)}_{0}(\B{r},\B{r},\omega)+\B{G}^{(2)}_{\delta}(\B{r},\B{r},\omega) .\label{SecondOrderGFBasic}
\end{equation}
Running through the analysis in this section again but with the two-sided curl $\nabla \times  \B{G}^{(2)}(\B{r},\B{r}',\omega) \times \overleftarrow{\nabla}' $ yields 
\begin{align}
\nabla & \times  \B{G}^{(2)}(\B{r},\B{r}',\omega)  \times \overleftarrow{\nabla}' \notag \\
&=\nabla \times \left[ \B{G}^{(2)}_{0}(\B{r},\B{r}',\omega)+\B{G}^{(2)}_{\delta}(\B{r},\B{r}',\omega) \right] \times \overleftarrow{\nabla}',
\end{align}
so we can definit electric and magnetic-type quantities as
\begin{align}
\B{G}^{(2)}_{\mathcal{E}}(\B{r},\B{r},\omega) &= \B{G}^{(2)}(\B{r},\B{r},\omega) \label{Electric0}\\
\B{G}^{(2)}_{\mathcal{B}}(\B{r},\B{r},\omega) &=\lim_{\B{r}\to \B{r}'}\nabla \times  \B{G}^{(2)}(\B{r},\B{r}',\omega) \times \overleftarrow{\nabla}' \label{Magnetic0},
\end{align}
which are precisely the quantities required to work out the Casimir energy density \eqref{EnergyStartPoint}. 

For $\B{r}\notin V$, the first term of Eq.~\eqref{SecondOrderGFBasic} is identical to \eqref{GSecondOrderBetweenPlates}, so the calculation shown in section \ref{SecondOrderForceSection} may be followed exactly with result given by eq.~\eqref{BasicSOGFBetweenPlates};
\begin{equation}
\B{G}^{(2)}_{0}(\B{r}\notin V,\omega)=\B{G}^{(2)}_\text{prop}(\B{r},\B{r},\omega).
\end{equation}
At this point we switch to reporting only the trace of the various Green's functions, which is what we require for the energy. We do this because of a proliferation of terms which simplify considerably once the trace is taken. 

\subsubsection{Second order: Electric terms}

Beginning with the electric-type term given by Eq.~\eqref{Electric0} for the central region $0<z<L$ we find:
\begin{align}
&\text{tr}\left[ \B{G}^{(2)}_{\mathcal{E}}(0<z<L,i\xi)\right]\notag \\ &=\frac{1}{16 \pi \xi^2   \epsMid^3(i\xi) } \int_0^\infty  \!\! dk_\parallel k_\parallel \frac{S^\mathcal{E}_\text{mid}(k_\parallel,\xi)
}{\kappa^5(k_\parallel,\xi)}f^\mathcal{E}_\text{mid}(k_\parallel, \xi) \label{SecondOrderElectricTraceCenter}, \end{align}
where 
\begin{align}
f^\mathcal{E}_\text{mid}(k_\parallel, \xi)  = 2 k_\parallel^6 +& 5 \xi ^2  k_\parallel^4 \epsMid(i\xi) \notag  \\& +3k_\parallel^2 \epsMid^2(i\xi) \xi^4 + \epsMid^3(i\xi) \xi^6,
\end{align}
and
\begin{align}
 S^\mathcal{E}_\text{mid}(\xi,  k_\parallel)=&e^{-2 z \kappa(k_\parallel,\xi)}+e^{2(z-L) \kappa(k_\parallel,\xi)}\notag \\
&+ \Big\{ f_\text{mid}^{-1}(k_\parallel, \xi) \xi ^2 \epsMid(i\xi)e^{-2 L \kappa(k_\parallel,\xi)} \notag \\
&\times [  2 k_\parallel^4+2 \xi ^2 k_\parallel^2 \epsMid(i\xi)+\xi ^4 \epsMid^2(i\xi) ] \Big\}. 
\end{align}
For the left region $z<0$, we find
\begin{align}
&\text{tr}\left[ \B{G}^{(2)}_{\mathcal{E}}(z<0,i\xi)\right]=\frac{1}{16 \pi   \epsMid^3(i\xi) \xi^2 }\int_0^\infty  dk_\parallel k_\parallel \notag \\  & \times \frac{S_\text{left}(k_\parallel,\xi)}{\kappa^5(k_\parallel,\xi)}f^\mathcal{E}_\text{left} (k_\parallel, \xi), \label{SecondOrderElectricTraceLeft}\end{align}
with
\begin{align}
f^\mathcal{E}_\text{left}(k_\parallel, \xi)=& -6 k_\parallel^6+2 \xi ^6 \epsMid^3(i\xi) \left[z\kappa(k_\parallel,\xi)-1\right] \notag \\
& +\xi ^4 k_\parallel^2 \epsMid^2(i\xi) \left[4 z\kappa(k_\parallel,\xi)-7\right]\notag \\
&+\xi ^2 k_\parallel^4 \epsMid(i\xi) \left[4 z\kappa(k_\parallel,\xi)-13\right],
\end{align}
and 
\begin{align}
S^\mathcal{E}_\text{left}(k_\parallel,\xi)= e^{2 z\kappa(k_\parallel,\xi)}-e^{-2 (L-z)\kappa(k_\parallel,\xi)}, \label{SLeft}
\end{align}
with the corresponding quantity for the right slab being obtained as usual by taking $z \to L-z$, i.e.:

\begin{equation}
 \B{G}^{(2)}_{\mathcal{E}}(z>L,i\xi) =\B{G}^{(2)}_{\mathcal{E}}(z<0,i\xi)\big|_{z\to L-z} \; . 
\end{equation}

\subsubsection{Second order: Magnetic terms}

Similarly for the magnetic-type term given by Eq.~\eqref{Magnetic0} we find:
\begin{align}
&\text{tr}\left[ \B{G}^{(2)}_{\mathcal{B}}(0<z<L,i\xi)\right]\notag \\ &=\frac{1}{16 \pi \xi^2   \epsMid^3(\xi) } \int_0^\infty  dk_\parallel k_\parallel \frac{S^\mathcal{B}_\text{mid}(k_\parallel,\xi)
}{\kappa^5(k_\parallel,\xi)}f^\mathcal{B}_\text{mid}(k_\parallel, \xi), \label{SecondOrderMagneticTraceCenter} \end{align}
where
\begin{equation}
f^\mathcal{B}_\text{mid}(k_\parallel, \xi) =\xi ^4\epsMid(i \xi) \left(k_\parallel^4+3 \xi ^2 k_\parallel^2 \epsMid(i \xi)+\xi ^4\epsMid(i \xi)\right),
\end{equation}
and
\begin{align}
& S^\mathcal{B}_\text{mid}(k_\parallel,\xi)=e^{-2z \kappa(k_\parallel,\xi)}+e^{2 (z-L) \kappa(k_\parallel,\xi)}\notag \\
&-\Big \{\xi ^4\epsMid(i \xi)\left[f^{\mathcal{B}}_\text{mid}(k_\parallel,\xi)\right]^{-1} \notag \\
&\times {\left[2 k_\parallel^4+2 \xi ^2 k_\parallel^2 \epsMid(i \xi)+\xi ^4\epsMid^2(i \xi)\right] e^{-2L \kappa(k_\parallel,\xi)}}\Big\},
\end{align}

Finally for the left region ($z<0$), we find
\begin{align}
&\text{tr}\left[ \B{G}^{(2)}_{\mathcal{B}}(z<0,i\xi)\right]\notag \\ &\!\!=\frac{1}{16 \pi \xi^2   \epsMid^3(i\xi) } \int_0^\infty  dk_\parallel k_\parallel \frac{S^\mathcal{B}_\text{left}(k_\parallel,\xi)
}{\kappa^5(k_\parallel,\xi)}f^\mathcal{B}_\text{left}(k_\parallel, \xi) , \label{SecondOrderMagneticTraceLeft} \end{align}
with
\begin{align}
f^\mathcal{B}_\text{left}(k_\parallel, \xi) =& \, \xi ^4 \epsMid^2(i \xi)  \Big\{k_\parallel^4+ 2 \xi ^4 z \epsMid^2 (i \xi)\kappa(k_\parallel,\xi)\notag \\
&+\xi ^2 k_\parallel^2 \epsMid(i \xi) \left[4 z\kappa(k_\parallel,\xi)-1\right]\Big\},
\end{align}
and where $S^\mathcal{B}_\text{left}(k_\parallel,\xi)$ is identical to that for the electric-type terms discussed in the previous section
\begin{equation}
S^\mathcal{B}_\text{left}(k_\parallel,\xi)= S^\mathcal{E}_\text{left}(k_\parallel,\xi).
\end{equation}
Finally we again note that the corresponding result $ \B{G}^{(2)}_{\mathcal{B}}(z>L,i\xi)$ for the right-hand slab is given by
\begin{equation}
 \B{G}^{(2)}_{\mathcal{B}}(z>L,i\xi) =\B{G}^{(2)}_{\mathcal{B}}(z<0,i\xi)\big|_{z\to L-z} \; . 
\end{equation}

\subsubsection{Second order: Total energy}

We now have all the ingredients required to work out the energy density in all three regions. Beginning with the central region, we need to substitute Eqs.~\eqref{SecondOrderElectricTraceCenter} and \eqref{SecondOrderMagneticTraceCenter} into Eq.~\eqref{SecondOrderEnergyInCenter}

\begin{align}
&\langle \rho_\text{mid} \rangle^{(2)}\equiv \langle \rho(0<z<L) \rangle ^{(2)} =\frac{1}{32\pi^2}\int_0^\infty \frac{d\xi}{\xi^2 }  \int_0^\infty  dk_\parallel k_\parallel \notag \\
&\times \frac{[\delta \epsilon(i \xi)]^2}{ \kappa^5(k_\parallel,\xi)\epsMid^3(i \xi) }   \Big\{  -\xi^2\epsMid(i\xi){S^\mathcal{E}_\text{mid}(k_\parallel,\xi)}f^\mathcal{E}_\text{mid}(k_\parallel, \xi)\notag \\
&\qquad \qquad\qquad \qquad\quad +{S^\mathcal{B}_\text{mid}(k_\parallel,\xi)}f^\mathcal{B}_\text{mid}(k_\parallel, \xi)\Big\}. \label{SecondOrderEnergyInCenterNotEvaluated}
\end{align}
Equation \eqref{SecondOrderEnergyInCenterNotEvaluated} is valid for arbitrary dispersive media. However, in order to facilitate comparison with known analytic results we now restrict ourselves to non-dispersive media $\epsilon(\omega) \to \epsilon$, $\epsilon_\text{mid}(\omega) \to \epsilon_\text{mid}$. Changing variables to a polar co-ordinate system defined by $k_\parallel = x \cos \theta$, $\xi = (x/\sqrt{\epsMid}) \sin \theta$, we find after a considerable amount of algebra and integration over $\theta$;
\begin{align}
\langle &\rho(0<z<L)  \rangle ^{(2)} =-\frac{[\epsilon-\epsilon_\text{mid}]^2}{1680 \pi ^2\epsMid^{5/2}}\int_0^\infty dx \, x^3 \notag \\
&\times \Big[ 64( e^{-2 z x}+e^{2 (z-L) x} )+43 e^{-2L x} \Big] \notag  \\
&\qquad =- \frac{[\epsilon-\epsilon_\text{mid}]^2}{{4480 \pi ^2\epsilon_\text{mid}^{5/2}}}\left[\frac{43}{L^4}+\frac{64}{(L-z)^4}+\frac{64}{z^4}\right] . \label{SecondOrderEnergyInCenterEvaluated}
\end{align}
Moving on to the region within the slabs $\B{r} \in \CasVol$, we have Eq.~\eqref{SecondOrderEnergyInSlabs} for the energy density in this region
\begin{align}
&\langle \rho(\B{r}\in \CasVol) \rangle ^{(2)} =\frac{1}{2\pi}\lim_{\B{r} \to \B{r}'} \int_0^\infty d\xi [\delta \epsilon(i\xi)]^2 \notag \\
& \times \text{tr} \Big\{-\xi^2\left[ \B{G}^{(1)}_\mathcal{E}(\B{r},\B{r}',\omega)+ \epsMid(i\xi)  \B{G}^{(2)}_\mathcal{E}(\B{r},\B{r}',\omega)\right]  \notag \\
&\qquad \qquad\qquad \qquad\qquad \qquad +  \B{G}^{(2)}_\mathcal{B}(\B{r},\B{r}',\omega)\Big\} ,
\end{align}
we see that we also need the trace of the first-order Green's function of the electric type in order to work out the second-order contribution to the energy. For convenience we write this in the same form as the traces of the second-order Green's functions [eqs.~\eqref{SecondOrderElectricTraceCenter}, \eqref{SecondOrderElectricTraceLeft}, \eqref{SecondOrderMagneticTraceCenter} and \eqref{SecondOrderMagneticTraceLeft}], giving
\begin{align}
&\text{tr}\left[ \B{G}^{(1)}_{\mathcal{E}}(z<0,i\xi)\right]=\frac{1}{16 \pi   \epsMid^3(i\xi) \xi^2 }\int_0^\infty  dk_\parallel k_\parallel \notag \\  & \times \frac{S_\text{left}(k_\parallel,\xi)}{\kappa^5(k_\parallel,\xi)}f^{\mathcal{E},(1)}_\text{left} (k_\parallel, \xi)\label{FirstOrderElectricTraceLeft} ,\end{align}
where
\begin{align}
f^{\mathcal{E},(1)}_\text{left} (k_\parallel, \xi)&= 2\epsMid  (k_\parallel,\xi) [k_\parallel^2+\xi ^2 \epsMid(k_\parallel,\xi)] \notag \\
&\times [2 k_\parallel^4+2 \xi ^2 k_\parallel^2 \epsMid(i\xi)+\xi ^4 \epsMid^2(i\xi)],
\end{align}
and where $S_\text{left}$ is given by Eq.~\eqref{SLeft}. Combining eqs.~\eqref{SecondOrderElectricTraceLeft}, \eqref{SecondOrderMagneticTraceLeft} and \eqref{FirstOrderElectricTraceLeft} we have:
\begin{align}
&\langle \rho_\text{left} \rangle^{(2)}\equiv \langle \rho(z<0) \rangle ^{(2)} =\frac{1}{32\pi^2}\int_0^\infty \frac{d\xi}{\xi^2 }  \int_0^\infty  dk_\parallel k_\parallel \notag \\
&\times \frac{[\delta \epsilon(i\xi)]^2}{ \kappa^5(k_\parallel,\xi)\epsMid^3(i\xi) }   \Big\{  -\xi^2{S^\mathcal{E}_\text{left}(k_\parallel,\xi)}\Big[ \epsMid(i\xi)f^\mathcal{E}_\text{left}(k_\parallel, \xi)\notag \\ &+f^{\mathcal{E},(1)}_\text{left} (k_\parallel, \xi)\Big]+{S^\mathcal{B}_\text{left}(k_\parallel,\xi)}f^\mathcal{B}_\text{left}(k_\parallel, \xi)\Big\} \label{SecondOrderEnergyLeftEvaluated} \; . 
\end{align}
Following our previous approach of restricting ourselves to non-dispersive media $\epsilon(\omega) \to \epsilon$, $\epsilon_\text{mid}(\omega) \to \epsilon_\text{mid}$, we find for the final result for the energy density in the left slab given by Eq.~\eqref{SecondOrderEnergyLeftEvaluated} becomes
\begin{equation}
\langle \rho_\text{left} \rangle^{(2)}=\frac{[\epsilon-\epsilon_\text{mid}]^2}{560\pi^2 \epsMid^{5/2}}\left[ \frac{13 z}{(L-z)^5}-\frac{9 L}{(L-z)^5}+\frac{13}{z^4} \right] ,
\end{equation}
which was derived by following by identical steps to those which took Eq.~\eqref{SecondOrderEnergyInCenterNotEvaluated} to Eq.~\eqref{SecondOrderEnergyInCenterEvaluated}. As usual we find the corresponding quantity in the right-hand slab by letting $z \to L-z$
\begin{align}
\langle \rho_\text{right} \rangle^{(2)}\equiv  \langle \rho(z>L) \rangle ^{(2)}  =&\frac{[\epsilon-\epsilon_\text{mid}]^2}{560\pi^2 \epsMid^{5/2}} \Bigg[ \frac{13 (L-z)}{z^5}\notag \\
& -\frac{9 L}{z^5}+\frac{13}{(L-z)^4} \Bigg] \label{SecondOrderEnergyRightSlab},
\end{align}
so that the total energy is given by
\begin{align}
E^{(2)}_\text{Cas} = \int_{-\infty}^0 dz \langle \rho_\text{left} \rangle^{(2)}+&\int_0^L dz\langle \rho_\text{mid} \rangle^{(2)}\notag \\
&+\int_L^\infty dz\langle \rho_\text{right} \rangle^{(2)} \; . 
\end{align}
In evaluating the above one finds that the integrals diverge at $z=0$ and $z=L$. Thus we introduce a small positive parameter $\delta$, and work instead with,
\begin{align}
 E^{(2),\delta}_\text{Cas} = \lim_{\delta \to 0} \Bigg[ \int_{-\infty}^{-\delta} dz\langle \rho_\text{left}\rangle^{(2)}&+\int_\delta^{L-\delta} dz\langle\rho_\text{mid}\rangle^{(2)}\notag \\
&\!\!\!\!\!\!\!\!\!\!\!\!\!\!\!\!\!\!\!\!+\int_{L+\delta}^\infty dz\langle\rho_\text{right}\rangle^{(2)} \Bigg] \label{EDelta}
\end{align}
resulting in,
\begin{equation}
E^{(2),\delta}_\text{Cas}= \frac{[\epsilon-\epsilon_\text{mid}]^2}{ \pi^2 \epsMid^{5/2}}\lim_{\delta \to 0} \left[ \frac{1}{168\delta^3 }-\frac{23}{1920}\frac{1}{L^3} +\mathcal{O}(\delta) \right],
\end{equation}
which is obviously divergent. The problematic term term is independent of $L$ so cannot contribute to the Casimir force, so from a physical standpoint there is no issue with the above equation. However, it seems strange that after all the subtractions of homogenous parts that one is still left with a term independent of $L$. This is because the divergent term represents the contribution from a surface charge density, arising because the dielectrics are not perfectly conducting. This part is immune from subtraction as a homogenous part, so appears in our final result for the Casimir energy. 
%
% This turns out to be simply a consequence of the choice of regulator $\delta$, which we implicitly chose in defining Eq.~\eqref{EDelta} to be the same in all three regions. For example, if one notes that close to the plates the energy densities diverge quartically and have ratios given by
%%
%\begin{align}
%\lim_{z \to 0} \left[  \langle \rho_\text{left}\rangle^{(2)}/\langle \rho_\text{mid}\rangle^{(2)}\right] &= -13/8 \notag \\
%\lim_{z \to L} \left[ \langle \rho_\text{right}\rangle^{(2)}/\langle \rho_\text{mid}\rangle^{(2)} \right] &= -13/8
%\end{align}
%%
%one is led to choosing the following regularization \footnote{The regulator in the regions $z<0$ and $z>L$ may be chosen to be the same since the integrals over those two regions are in fact identical if one makes the change of variables $z \to L-z$ in the $z>L$ integral.}
%%
%\begin{align}
%E^{(2)}_\text{Cas} = \lim_{\delta_1 \to 0} \Bigg[ & \int_{-\infty}^{-\delta_1}dz  \langle \rho_\text{left}\rangle^{(2)}+\int_{L+\delta_1}^\infty dz \langle \rho_\text{right}\rangle^{(2)}\notag \Bigg] \notag \\
%&+ \lim_{\delta_2 \to 0}  \int_{\delta_2 \sqrt[3]{8/13}}^{L-{\delta_2\sqrt[3]{8/13}}} dz \langle \rho_\text{mid}\rangle^{(2)}
%\end{align}
%%
%which leads directly to
%%
%\begin{equation}
%E^{(2)}_\text{Cas}=-\frac{23[\epsilon-\epsilon_\text{mid}]^2}{1920 \pi^2 \epsMid^{5/2}}\frac{1}{ L^3}  \; . 
%\end{equation}
%
The force is then found by taking the (negative) derivative of the this with respect to the plate separation $L$, so the divergent term disappears, allowing us to take the $\delta \to 0 $ limit and finally find
\begin{equation}
F^{(2)}_\text{Cas}=-\frac{dE^{(2)}_\text{Cas}}{dL}=-\frac{23[\epsilon-\epsilon_\text{mid}]^2}{640 \pi^2 \epsMid^{5/2}}\frac{1}{ L^4} ,
\end{equation}
in agreement with \eqref{ForceSecondOrderFinalResult}, and consequently also in agreement with \cite{SchwingerCasimir}. This completes our demonstration of how dielectric-contrast perturbation theory may be used to find the Casimir energy density for some arrangement of objects. 

\section{Numerical example}\label{NumericalSection}

As discussed in section \ref{Intro}, the real power of a dielectric-contrast perturbation theory approach to the Casimir force lies in its application to geometries for which the exact Green's function is not known. In this section we give a short example of the application of this method to such a system.

\begin{figure}[h!]
\includegraphics[width=\columnwidth]{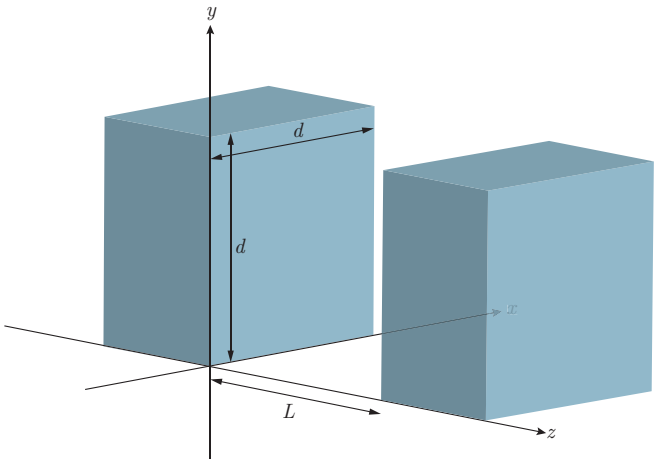}
\caption{(Color online) Geometry $\BoxVol$. Each object has infinite extent in the $z$ direction moving away from the gap.} \label{NumGeom}
\end{figure}

To begin with, we note that up to Eq.~\eqref{GijFirstOrderMiddle}, none of the expressions in this work make any reference to the actual shape $\CasVol$ of the Casimir geometry, apart from the specification of wether $\B{r}$ is inside or outside the objects. Thus we may in fact take $\CasVol \to \B{V}$, with $\B{V}$ being a general volume describing the objects.  Taking the expression \eqref{FirstOrderEnergyInCenter} for the first-order contribution to the energy density outside the objects
\begin{align}
\langle  \rho&(\B{r}\notin \B{V})  \rangle^{(1)}  =\frac{1}{2\pi} \int_0^\infty d\xi  \delta \epsilon(\omega) \notag \\
& \times  \text{tr} \Big[-\xi^2 \epsMid(\omega)  \B{G}^{(1)}_\mathcal{E}(\B{r},\B{r},\omega)+  \B{G}^{(1)}_\mathcal{B}(\B{r},\B{r},\omega)\Big], \label{RhoForNumerics}
\end{align}
and the expression \eqref{GijFirstOrderMiddle} for the first-order contribution to the scattering Green's function in that region for a general geometry $\B{V}$,
\begin{align}
\B{G}_{ij}^{(1)}(\B{r},\B{r},\omega)&=\omega^2 [\delta \epsilon(\omega)]   \int_\B{V} d^3 \B{s}\notag \\
&\times \B{H}^{(0)}_{\text{prop},ik}(\B{r},\B{s},\omega)\B{H}^{(0)}_{\text{prop},kj}(\B{s},\B{r},\omega ),\end{align}
we may now choose $\B{V}$ to describe any geometry for which we would like to calculate the Casimir energy density. We choose a box-like geometry $\BoxVol$ which is identical to $\CasVol$ considered previously in this work, but with its extent in the $x$ and $y$ directions limited such that its cross section is a square of area $d^2$, as shown in fig.~\ref{NumGeom}. The two objects remain infinite in the $z$ direction, and their separation is still $L$. The gap between will be taken to consist of vacuum, and the boxes are taken as non-dispersive with dielectric function $\epsilon$. 

For comparison's sake we quote the first-order term in the energy density in the gap between the slabs for the infinite case ($\CasVol$) with $\epsMid(\omega) \to 1$ and $\epsilon(\omega)\to \epsilon$, given by
\begin{equation}
\langle  \rho^\text{C}_\text{mid}  \rangle^{(1)}\equiv \langle  \rho (\B{r}\notin \CasVol)  \rangle^{(1)} = \frac{\epsilon-1}{40\pi^2} \left[\frac{1}{(L-z)^4}+\frac{1}{z^4} \right],
\end{equation}
which is found by carrying out the integral shown in \eqref{FirstOrderEnergyDensityCenter}. This also trivially coincides with the proximity-force approximation (PFA) result for this geometry. 

\begin{figure}
\includegraphics[width=\columnwidth]{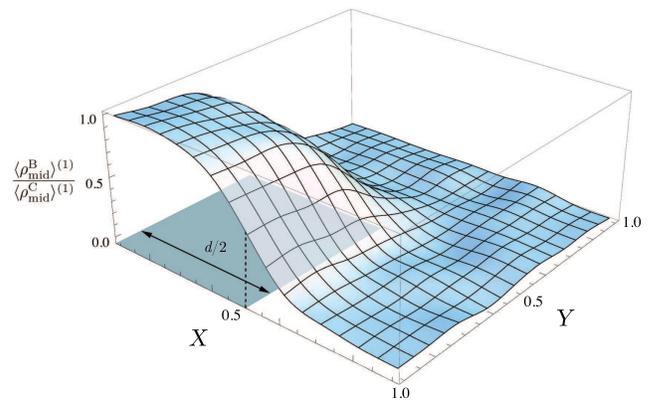}
\caption{(Color online) Energy density in units of the infinite plate energy $\langle  \rho^\text{C}_\text{mid}  \rangle^{(1)}$ for the box geometry $\BoxVol$ with $\lambda = 2$, $\epsilon=2$ and $Z=1/5$, plotted against the dimensionless variables $X$ and $Y$ [c.f. Eq.~\eqref{DimVars}]. For clarity we have only shown the region $X>0, Y>0$, but the other three quadrants have identical results because of the symmetry of the physical setup. Since we are plotting in units of the infinite plate energy $\langle  \rho^\text{C}_\text{mid}  \rangle^{(1)}$ we do not need to supply explicit values for $L$ or $d$, only their ratio $\lambda$.}
 \label{EnergyGraph}
\end{figure}

\begin{figure}[h!]
\includegraphics[width=\columnwidth]{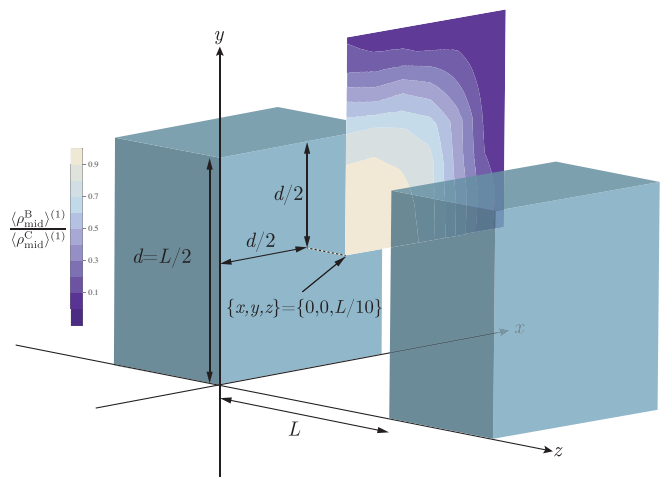}
\caption{(Color online) Alternative representation of the same results as those shown in fig.~\ref{EnergyGraph}.}\label{BlocksWithGraph}
\end{figure}

Using the Monte Carlo integration routines of \emph{Mathematica}, we evaluate Eq.~\eqref{RhoForNumerics} numerically with
\begin{align}
\int_\B{V} d^3 \B{s}\to \int_\BoxVol d^3 \B{s} = &\int_{-d/2}^{d/2}  ds_x  \int_{-d/2}^{d/2} ds_y \notag \\
& \times \left(\int_{-\infty}^0 ds_z + \int_L^\infty ds_z \right) \, .
\end{align}
 We find the results shown in figs \ref{EnergyGraph} and \ref{BlocksWithGraph}, where we plot the energy density $\langle  \rho^\text{B}_\text{mid}  \rangle^{(1)}$ near one of the objects in units of the infinite plate energy $\langle  \rho^\text{C}_\text{mid}  \rangle^{(1)}$ in terms of dimensionless parameters defined by
 \begin{align}
\lambda &\equiv L/d & X  &\equiv x/d & Y& \equiv y/d & Z\equiv z/d \; .  \label{DimVars}
\end{align}
The results for the box geometry $\BoxVol$ near $\{x,y\}=\{0,0\}$ are similar to those for the Casimir geometry $\CasVol$. This is expected because the centre of each plate is obviously the furthest point from the edges, so is the point at which finite-size effects are least important. We also show qualitative agreement with the world-line approach of \cite{GiesKlingmullerEdgeEffects}, where it was noted that the Casimir energy for a scalar field under perfectly reflecting boundary conditions has a peak which extends into an `outside' region (in our example this corresponds to $|x|,|y|>d/2$), our results also have this property [c.f. dashed line in fig \ref{EnergyGraph}]. The physical interpretation of this in \cite{GiesKlingmullerEdgeEffects} was that a worldline can intersect both plates even if its center of mass (see, for example \cite{Gies2001353}) is in the outside region -- in our language this is a consequence of it being possible that the observation point $\B{r}$ is in the outside region while at least one scattering event is within the slabs.

\section{Conclusions}

We have investigated the Casimir force between parallel infinite dielectric slabs using dielectric contrast perturbation theory within the canonically-quantized framework of macroscopic QED, as opposed to the often-used (arguably classical) original theory of Lifshitz. We have shown that the method reproduces previous results, and noted some important properties of the Born series upon its application to the Casimir effect -- in particular some subtleties when working with the Green's function in a region whose dielectric function is not equal to the unperturbed one. We have also demonstrated the physically intuitive fact that one will always need to consider at least two scattering events to find Casimir forces in a multiple-scattering approach. In contrast to this, we have observed the completely \emph{unintuitive} fact that the second-order terms in the Born series of the Green's function must be taken into account even when calculating the \emph{first}-order Casimir energy.  Finally, to emphasise the power of the method, we have used a general expression of the Casimir energy to relatively easily find numerical results for a simple but non-trivial geometry. Application to more complex geometries does not represent too much of a challenge based on the results we have presented here, especially as nothing in this work constrains the unperturbed Green's function to be homogenous -- one can use any object for which the Green's function is analytically known (plane, cylinder, sphere) as the unperturbed geometry. This extension would have clear applications in studies of surface roughness and defects that affect the results of Casimir force experiments, as well as the development of future MEMS/NEMS technologies. 

\section{Acknowledgements}

It is a pleasure to thank Almut Beige for advice. The author also wishes to thank the referee for helpful comments and advice. Financial support from the UK Engineering and Physical Sciences Research Council (EPSRC) is gratefully acknowledged. 

\appendix

\section{Interplay between the unperturbed Green's function and the homogenous part of the whole Green's function}\label{ExampleAppendix}

Consider a region whose dielectric function is given by
\begin{equation}
\epsilon(\B{r}, \omega) = \epsilon^{(0)}(\omega)+\epsilon_p(\B{r}, \omega)
\end{equation}
with $\epsilon_p(\B{r}, \omega)$ representing a perturbation to a homogenous dielectric function $\epsilon^{(0)}(\omega)$. The total Green's function $\boldsymbol{\Gamma}\equiv \boldsymbol{\Gamma}(\B{r}, \B{r}',\omega)$ for the region is then given by
\begin{equation}
\boldsymbol{\Gamma} = \boldsymbol{\Gamma}^{(0)}+\boldsymbol{\Sigma}_p \label{TotalGFSymbolic}
\end{equation}
where $\boldsymbol{\Gamma}^{(0)}\equiv \boldsymbol{\Gamma}^{(0)} (\B{r}, \B{r}',\omega)$ is the Green's function for a region with dielectric function $\epsilon^{(0)}(\omega)$, and $\boldsymbol{\Sigma}_p$ represents the $n\geq 1$ terms of the Born series \eqref{ScatteringDerivation}. Taking the homogenous part of both sides of Eq~\eqref{TotalGFSymbolic} and rearranging we have
\begin{equation}
\B{F}[\boldsymbol{\Sigma}_p] = \B{F}\left[\boldsymbol{\Gamma}\right]-\boldsymbol{\Gamma}^{(0)}
\end{equation}
showing that choosing $\B{r}$ and $\B{r}'$ such that $\B{F}\left[\boldsymbol{\Gamma}\right]\neq \boldsymbol{\Gamma}^{(0)}$ will result in a non-zero homogenous part for the perturbation $\boldsymbol{\Sigma}_p$. We emphasize that even though the exact Green's function $\boldsymbol{\Gamma}$ is, in general, unknown, its homogenous part at a particular point $\B{r}=\B{r}_0$ \emph{is} always known since it may be obtained simply by finding the Green's function for a homogenous medium with permittivity $\epsilon(\omega) = \epsilon(\B{r}_0,\omega)$ -- this Green's function is analytically obtainable and well-known. 

Finally we note that since $\epsilon^{(0)}(\omega)$ is homogenous, we have $\B{F}[ \boldsymbol{\Gamma}^{(0)}]= \boldsymbol{\Gamma}^{(0)}$.  This means we have for the scattering Green's function $\B{G}=\boldsymbol{\Gamma}-\B{F}\left[\boldsymbol{\Gamma}\right]$;
\begin{equation}
\B{G} = \boldsymbol{\Sigma}_p-\B{F}[\boldsymbol{\Sigma}_p]
\end{equation}
showing that in regions with $\B{F}\left[\boldsymbol{\Gamma}\right]\neq \boldsymbol{\Gamma}^{(0)}$ one has to subtract a homogenous part after the evaluation of $\boldsymbol{\Sigma}_p$. It follows that
\begin{equation}
\B{G} = \boldsymbol{\Sigma}_p \qquad  \text{ if } \qquad \B{F}\left[\boldsymbol{\Gamma}\right]= \boldsymbol{\Gamma}^{(0)}
\end{equation}
The calculations of  \cite{buhmann2006born, GolestanianCasimir} were restricted to regions where the above condition holds, but we do not restrict ourselves to such situations here.

\section{Duality relations}\label{DualityAppendix}

It is well-known that $\B{G}$ and $\nabla \times \B{G} \times \overleftarrow{\nabla}'$ can be simply related to each other by using a duality relation $\epsilon \leftrightarrow \mu$ \cite{BuhmannScheelDuality}. For example, for a Green's function $\B{G}(\B{r},\B{r}',\omega )$ at $\B{r}\neq \B{r}'$ in a region with permittivity $\epsilon(\B{r})$ and permeability $\mu(\B{r})$ we have \cite{BuhmannScheelDuality}:
\begin{equation}
 \omega^2 \B{G}(\B{r},\B{r}' )|_{\epsilon \leftrightarrow \mu}=\frac{1}{\mu(\B{r})}\nabla \times \B{G}(\B{r},\B{r}' ) \times \overleftarrow{\nabla} \frac{1}{\mu(\B{r}')}
\end{equation}
This strongly suggest that we should be able to avoid the complication of actually having to take the two-sided curl in our Born series calculation. However, this is not the case. To see this, take as an example the $\B{M}$ and $\B{N}$-dependence of the first order term in the Born series
\begin{align}
&\left[\B{M}(\B{r})\otimes \B{M}(\B{s})+\B{N}(\B{r})\otimes \B{N}(\B{s})\right]\notag \\
&\qquad \qquad \cdot \left[\B{M}(\B{s})\otimes \B{M}(\B{r}')+\B{N}(\B{s})\otimes \B{N}(\B{r})\right] \label{MNDep}
\end{align}
We can take the two-sided curl to find
\begin{align}
&k(\B{r}) k(\B{r}')\left[\B{N}(\B{r})\otimes \B{M}(\B{s})+\B{M}(\B{r})\otimes \B{N}(\B{s})\right]\notag \\
&\qquad \qquad \cdot \left[\B{M}(\B{s})\otimes \B{N}(\B{r}')+\B{N}(\B{s})\otimes \B{M}(\B{r})\right] \label{DualityProb}
\end{align}
Quite apart from the complication that $k (\B{r})=\sqrt{\epsilon(\B{r})}\omega $ is not, in general, equal to $k (\B{r}')$ in an inhomogenous system such as ours, this is not obviously related to Eq.~\eqref{MNDep} in any particularly simple way, meaning there is no gain in attempting to relate each term in the Born series of $\B{G}$ and $\nabla \times \B{G} \times \overleftarrow{\nabla}'$ in this calculation. However we emphasise that the duality relations in \cite{BuhmannScheelDuality} hold on sufficiently general grounds that this work must be reproducible using them, we have simply chosen not to since explicitly taking the two-sided curl proves to be not significantly more complicated than using a duality relation because of the kind of issue shown in Eq.~\eqref{DualityProb}.

\end{document}